\documentclass[12pt,psfig,dvips,preprint]{aastex}
\usepackage{amssymb}
\usepackage{amsmath}
\usepackage{graphicx}

\newcommand\HII{H\,{\sc ii}~}

\newcommand\kms{km~s$^{-1}$}

\newcommand\pp{$^{\prime\prime}$}
\newcommand\um{$\mu$m}

\newcommand\q{$\sim$}

\newcommand\h{H$_{2}$~}
\newcommand\msun{M$_{\odot}$}

\newcommand{\meth}{CH$_3$OH}

\newcommand{\vlsr}{$v_{LSR}$}

\begin{document}
\shortauthors{Cyganowski et al.}

\title{A Catalog of Extended Green Objects (EGOs) in the GLIMPSE Survey: A new
  sample of massive young stellar object outflow candidates}
\author{C.J. Cyganowski\altaffilmark{1}, B.A. Whitney\altaffilmark{2},
E. Holden\altaffilmark{1}, E. Braden\altaffilmark{1},
C.L. Brogan\altaffilmark{3}, E. Churchwell\altaffilmark{1}, R. Indebetouw\altaffilmark{3,4}, 
D. F. Watson\altaffilmark{5}, B. L. Babler\altaffilmark{1},
R. Benjamin\altaffilmark{6}, M. Gomez\altaffilmark{7}, M. R. Meade\altaffilmark{1},
M. S. Povich\altaffilmark{1}, T. P. Robitaille\altaffilmark{8}, C. Watson\altaffilmark{9}}

\email{ccyganow@astro.wisc.edu}

\altaffiltext{1}{University of Wisconsin, Madison, WI 53706}
\altaffiltext{2}{Space Science Institute, 4750 Walnut St. Suite 205, Boulder,
CO 80301, bwhitney@spacescience.org} 
\altaffiltext{3}{NRAO, 520 Edgemont Rd, Charlottesville, VA 22903}
\altaffiltext{4}{University of Virginia, Astronomy Dept., P.O. Box 3818,
  Charlottesville, VA, 22903-0818} 
\altaffiltext{5}{Department of Physics and Astronomy, Vanderbilt University, Nashville, TN 37235}
\altaffiltext{6}{Department of Physics, University of Wisconsin at Whitewater,
800 West Main Street, Whitewater, WI 53190} 
\altaffiltext{7}{Observatorio Astronomico, Universidad Nacional de Cordoba, Argentina, Laprida 854, 5000
Cordoba, Argentina} 
\altaffiltext{8}{SUPA, School of Physics and Astronomy, University of
St. Andrews, North Haugh, St. Andrews, UK} 
\altaffiltext{9}{Department of Physics, Manchester College, North Manchester, IN 46962}

\begin{abstract}
Using images from the \emph{Spitzer} GLIMPSE Legacy survey, we have identified
more than 300 extended 4.5 \um\/ sources (abbreviated EGO, Extended Green
Object, for the common coding of the [4.5] band as green in 3-color composite
IRAC images).  We present a catalog of these EGOs, including integrated flux density
measurements at 3.6, 4.5, 5.8, 8.0, and 24 \um\/ from the GLIMPSE and MIPSGAL
surveys.  The average angular separation between a source in our sample and
the nearest IRAS point source is $>$1\arcmin.  The majority of EGOs are
associated with infrared dark clouds (IRDCs), and where high-resolution 6.7
GHz \meth\/ maser surveys overlap the GLIMPSE coverage, EGOs and 6.7 GHz
\meth\/ masers are strongly correlated.  Extended 4.5 \um\/ emission is thought
to trace shocked molecular gas in protostellar outflows; the association of
EGOs with IRDCs and 6.7 GHz \meth\/ masers suggests that extended 4.5 \um\/
emission may pinpoint outflows specifically from massive protostars.  The
mid-infrared colors of EGOs lie in regions of color-color space occupied by
young protostars still embedded in infalling envelopes.  
\end{abstract}


\section{Introduction}

A fundamental limitation in the study of high-mass star formation is
the identification of massive young stellar objects (MYSOs),
particularly young MYSOs that have not yet formed hypercompact (HC) or
ultracompact (UC) \HII regions.\footnote{We define MYSOs to be young
stellar objects (YSOs) that will eventually become main-sequence O or
early B type stars (M$_{*}\gtrsim$ 8 \msun).}  The largest samples
assembled to date have been drawn from the \emph{IRAS} Point Source
Catalog \citep[e.g.][]{Sr02,Mol96}, and are limited by its poor
resolution (which often blends emission from multiple sources) and
positional uncertainty.  The \emph{MSX} Point Source Catalog, the
basis for the RMS sample \citep[e.g][]{Urquhart08,Hoare05,Hoare04},
represents a vast improvement, but emission from clustered sources may
still be blended at the resolution of \emph{MSX}. 

Recent large-scale, high-angular resolution \emph{Spitzer} surveys
of the Galactic plane using the InfraRed Array Camera \citep[IRAC, 3.6, 4.5,
5.8, and 8.0 \um, resolution $<$2\pp\/ in all bands;][]{Fazio04} and Multiband
Imaging Photometer for Spitzer \citep[MIPS, resolution \q6\pp\/ at
24 \um\/;][]{Rieke04} present an opportunity to compile a new, less-confused
sample of IR-selected MYSO candidates.  The IRAC 4.5 \um\/ band provides a
promising new diagnostic in the search for MYSO candidates.  Extended 4.5
\um\/ emission is a conspicuous and ubiquitous feature of \emph{Spitzer}
images of known massive star-forming regions (MSFRs), including DR21
\citep[][and references therein]{Davis07,Smith06}, S255N
\citep{2007AJ....134..346C}, NGC6334I(N) \citep{Hunter06},
G34.4+0.23 \citep{2007ApJ...669..464S}, and IRAS 18566+0408 \citep{Araya07}.
The IRAC 4.5 \um\/ band contains both \h\/ (v=0-0, S(9,10,11)) lines and CO
(v=1-0) bandheads \citep[c.f. Figure 1 of][]{Reach06}.  All of these lines may
be excited by shocks, such as those expected when protostellar outflows crash
into the ambient ISM.  Models of protostellar jets by \citet{SmithRosen}
predict that, of the IRAC bands, the integrated \h\/ line luminosity will be
greatest in band 2 (4.5 \um).  Observationally, the extended 4.5 \um\/ emission
of the DR21 outflow has been interpreted as tracing shocked \h\/ based on
comparisons with NIR \h\/ narrowband images and \emph{ISO} spectra
\citep[][and references therein]{Davis07,Smith06}.

Extended 4.5 \um\/ emission is of particular interest as a new MYSO diagnostic
because the selection effects are orthogonal to those of traditional methods
of identifying MYSO candidates (e.g. applying color criteria to point source
catalogs, as done for \emph{IRAS} by \citet{Sr02} and \citet{Mol96} and for
\emph{MSX} by \citet{Hoare05,Hoare04}).  Even at the resolution of
IRAC, the environments in which MYSOs form--clusters, infrared dark clouds,
and nebulous regions--can complicate point-source-based selection methods.
Identifying sources with extended 4.5 \um\/ emission targets a population with
ongoing outflow activity and active, rapid accretion--a short-lived, and
relatively poorly understood, stage of massive protostellar evolution
\citep[c.f][]{ZY07}.

We have compiled a catalog of
over 300 extended 4.5 \um\/ sources in the GLIMPSE survey, abbreviated EGO
(Extended Green Object, for the common coding of the [4.5] band as green in
3-color composite IRAC images).  We present this catalog, split into four
sub-catalogs  (see \S3).
We present integrated flux densities in the IRAC 4.5 \um\/ band for all catalogued
EGOs, and investigate the nature of these sources by examining their
correlations with infrared dark clouds (IRDCs) and 6.7 GHz Class II \meth\/
masers.  For those EGOs for which counterparts could be identified and
photometered in all four IRAC bands and the MIPs 24 \um\/ band, we present
five-band photometry (integrated flux densities at 3.6 \um\/, 4.5 \um\/, 5.8 \um\/,
8.0 \um\/, and 24 \um\/), and investigate the EGOs' colors.  In \S2
we describe the data and techniques used to compile the catalog, in \S3 we
present the catalog, and in \S4 we discuss the nature of EGOs.

\section{The Data}

\subsection{Identification of EGOs}

This paper catalogs EGOs identified in the GLIMPSE survey area
(10$\degr <$\emph{l}$<$ 65$\degr$ and 295$\degr <$\emph{l} $<$ 350
$\degr$, \emph{b}=$\pm$1$\degr$).  To identify candidate outflows, we
displayed the GLIMPSE $l \times b$=3$\degr \times$ 2$\degr$ (1\farcs2
per pixel) mosaics\footnote{available at
http://data.spitzer.caltech.edu/popular/glimpse/20070416\_enhanced\_v2/3.1x2.4\_mosaics/}
as 3-color images with the IRAC 3.6 \um\/ image in blue, 4.5 \um\/ in
green, and 8.0 \um\/ in red.  Each image was searched by visual
examination for extended green emission, indicating excess at 4.5 \um.
The source list compiled from this search was then reviewed by two
additional authors independently, and only candidates agreed upon by
all three authors are included in the catalog.  Images in each of the
IRAC bands (3.6, 4.5, 5.8, and 8.0 \um\/), in the MIPS 24 \um\/ band,
and a three-color composite (3.6 \um\/ in blue, 4.5 \um\/ in green, and 8.0
\um\/ in red) for all catalogued EGOs are available online; an
example is shown in Figure~\ref{table1_example}.

Most catalogued EGOs are bright and relatively compact, though more extended
than point sources (angular extents range from a few to $>$30\arcsec).  EGOs
are readily distinguishable by eye from supernovae remnants (SNR).  The IRAC
emission from SNR may be dominated by shock-excited warm molecular hydrogen
\citep{Neufeld08}, but SNR are generally larger, more diffuse, and more
structured than EGOs \citep{Reach06}.  EGOs are also distinguishable from
Planetary Nebulae (PNe), which appear red in these 3-color images and
generally exhibit spherical morphology at MIR wavelengths.  There are three
regimes in which our EGO catalog is likely to be incomplete: (1) extended
green emission near bright sources; (2) extended green emission in sources
with significant PAH emission; and (3) diffuse green emission extended
$\lesssim$10\arcsec.  Image artifacts associated with bright IRAC sources may
obscure or confuse extended 4.5 \um\/ emission associated with these sources.
Bright PAH emission (e.g. in a YSO envelope) may overwhelm extended 4.5 \um\/
emission from an outflow.  Slightly extended ($\lesssim$10\arcsec) 4.5 \um\/
emission may be confused with multiple nearby or blended point sources (and
vice versa); requiring agreement among three observers may exclude some
slightly extended sources.\footnote{Of 326 sources identified in the
  initial search, 24 (\q 7\%) were excluded because the three
  observers did not agree on their inclusion.}  This paper does not catalog {\it point-like}
sources that appear green in 3-color images for two reasons: (1) point sources
with distinctive [3.6]-[4.5] colors may be identified via color selection in
the GLIMPSE point source catalog and archive
\citep[e.g.][]{Ellingsen06,Ellingsen07}; and (2) due to the flat extinction
curve between 4.5 and 8.0\um\/ \citep{Remy05}, highly extincted stars could
potentially appear green in 3.6/4.5/8.0\um\/ color images
\citep[e.g.][]{Rathborne05}.

\subsection{Photometry}

Areal photometry was performed on all catalogued EGOs, using ds9 funtools
\footnote{Available at http://hea-www.harvard.edu/saord/funtools/ds9.html}.
Arbitrarily shaped polygon apertures were made to closely match the source
morphology in 4.5 \um\/ images.  The same polygonal aperture was used for
flux density measurements in all four IRAC bands.  

A separate aperture was made to measure the local background, the flux density was
integrated over the source, and the area-corrected background subtracted.  The
flux densities of EGOs at 24 \um\/ were also measured, since this wavelength, combined
with IRAC, is a good indicator of evolutionary stage in protostars
\citep{Robitaille06}.  We measured 24 \um\/ flux densities from mosaics of the Basic
Calibrated Data (BCD) from the MIPSGAL survey.  Larger, circular or elliptical apertures
were used for the 24 \um\/ photometry because the beamsize at 24 \um\/ is at
least three times larger than at IRAC wavelengths.

Most EGOs are located in areas of high and/or variable background
emission, and uncertainty in the determination of the background level
is a dominant contributor to the overall uncertainty in the measured
flux densities.  For each band, three measurements were made of the
source integrated flux density, using three different background
regions.  Care was taken in placing the background regions to avoid
bright stars and areas of exceptionally strong or weak emission.  The
average of the three measured source flux densities is reported in
Tables~\ref{fluxtable1}-\ref{fluxoutflowonly}, along with the standard
deviation of the three measurements.  The standard deviation provides
an estimate of the uncertainty in the reported source flux densities
resulting from uncertain background subtraction.  The uncertainty due
to background subtraction is generally greatest for the 8 \um\/ flux
densities, as diffuse background emission is strongest and most
spatially variable in this band.  The other major source of
uncertainty in the measured source flux densities is
observer-dependent aperture selection.  To quantify this effect, a
sample of 15 sources was photometered by two additional authors;
Figure~\ref{3obs_apertures} shows the IRAC and MIPS apertures selected
by the three different observers for the example source shown in
Figure~\ref{table1_example}.  The mean observer-dependent uncertainty
is estimated to be \q15\% in the IRAC 3.6, 4.5, and 5.8 \um\/ bands
and the MIPS 24 \um\/ band and about \q 25\% in the IRAC 8.0 \um\/
band, in addition to the background-dependent uncertainties listed in
Tables~\ref{fluxtable1}-\ref{fluxoutflowonly}.

\section{The Catalogs}

Tables~\ref{fluxtable1}-\ref{fluxtable3} present the four subcatalogs of EGOs:
``likely'' MYSO outflow candidates are listed in
Tables~\ref{fluxtable1}-\ref{fluxtable2}, ``possible'' MYSO outflow candidates
in Tables ~\ref{fluxtable1a}-\ref{fluxtable3}.  Table~\ref{fluxtable1}
presents integrated [3.6], [4.5], [5.8], [8.0], and [24] flux densities for ``likely''
MYSO outflow candidates with counterparts in all five bands (97 sources); for
``likely'' sources in which confusion and/or saturation are problematic in
some bands, Table~\ref{fluxtable2} presents integrated [4.5] flux densities only (36
sources).  Similarly, Table ~\ref{fluxtable1a} presents integrated [3.6], [4.5], [5.8], [8.0], and
[24] flux densities for ``possible'' MYSO outflow candidates with counterparts in all
five bands (75 sources); for ``possible'' sources in which confusion and/or
saturation are problematic in some bands, Table~\ref{fluxtable3} presents
integrated [4.5] flux densities only (90 sources).  Table~\ref{fluxoutflowonly}
reports integrated [3.6] and [4.5] flux densities, and [3.6-4.5] colors, for sources
in which extended outflow emission may be photometered independently of a 
central source (see discussion in \S4.3).

The sources in Tables~\ref{fluxtable1}-\ref{fluxoutflowonly} are organized by
increasing galactic longitude.  The columns in Tables~\ref{fluxtable1} and
\ref{fluxtable1a} are: (1) name (galactic coordinates), (2-3) position (RA and
Dec, rounded to the nearest 0.1 second and nearest arcsecond respectively),
(4) the area over which IRAC flux densities were integrated, in units of square
arcseconds, (5-8) integrated flux densities in the 3.6, 4.5, 5.8, and 8.0 \um\/ IRAC
bands (mJy), (9) the area over which the MIPS 24 \um\/ flux density was integrated, in
units of square arcseconds, (10) integrated flux density in the 24 \um\/ MIPS band
(mJy), (11) a flag for whether the source is saturated in any band and (12) a
list of bands in which the quoted flux density should be considered an upper limit
because the measurement may include flux from the psf of a nearby source.
Tables~\ref{fluxtable2} and ~\ref{fluxtable3} are similar in format, except
that only [4.5] flux densities are reported.

The coordinates in columns 2-3 in
Tables~\ref{fluxtable1}-\ref{fluxoutflowonly} are those of the
brightest 4.5 \um\/ emission associated with the EGO.  We emphasize
that the EGOs in Tables~\ref{fluxtable1}-\ref{fluxoutflowonly} have
angular extents on the sky ranging from a few to $>$30\pp, and many
are near other MIR GLIMPSE and/or MIPS sources.  To establish
correlations (or the lack thereof) between the EGOs presented here and
other datasets, or to plan followup observations, visual examination
of the GLIMPSE and MIPSGAL images should be used along with the listed
positions.
 
The errors quoted in Tables~\ref{fluxtable1}-\ref{fluxoutflowonly} are the
standard deviations of the three measurements made with different background
regions for each band (see \S2.2).  All reported flux densities are rounded to the
nearest 0.1 mJy, and the reported areas are rounded to the nearest square
arcsecond.

The categorization of an EGO as a ``likely'' or ``possible'' MYSO outflow
candidate was based primarily on the angular extent of the extended excess 4.5
\um\/ emission, e.g. the extent of diffuse green emission in 3-color images.
Three-color images of each EGO were reviewed by two of the authors; any source
in which either observer thought it was possible that multiple nearby point
sources and/or image artifacts from a bright IRAC source could be confused
with truly extended 4.5 \um\/ emission is considered a ``possible'' candidate.
These relatively compact sources are likely still good YSO candidates (see \S4), but not
necessarily MYSOs \emph{with outflows} and so likely to be actively accreting.

\begin{figure} 
\plotone{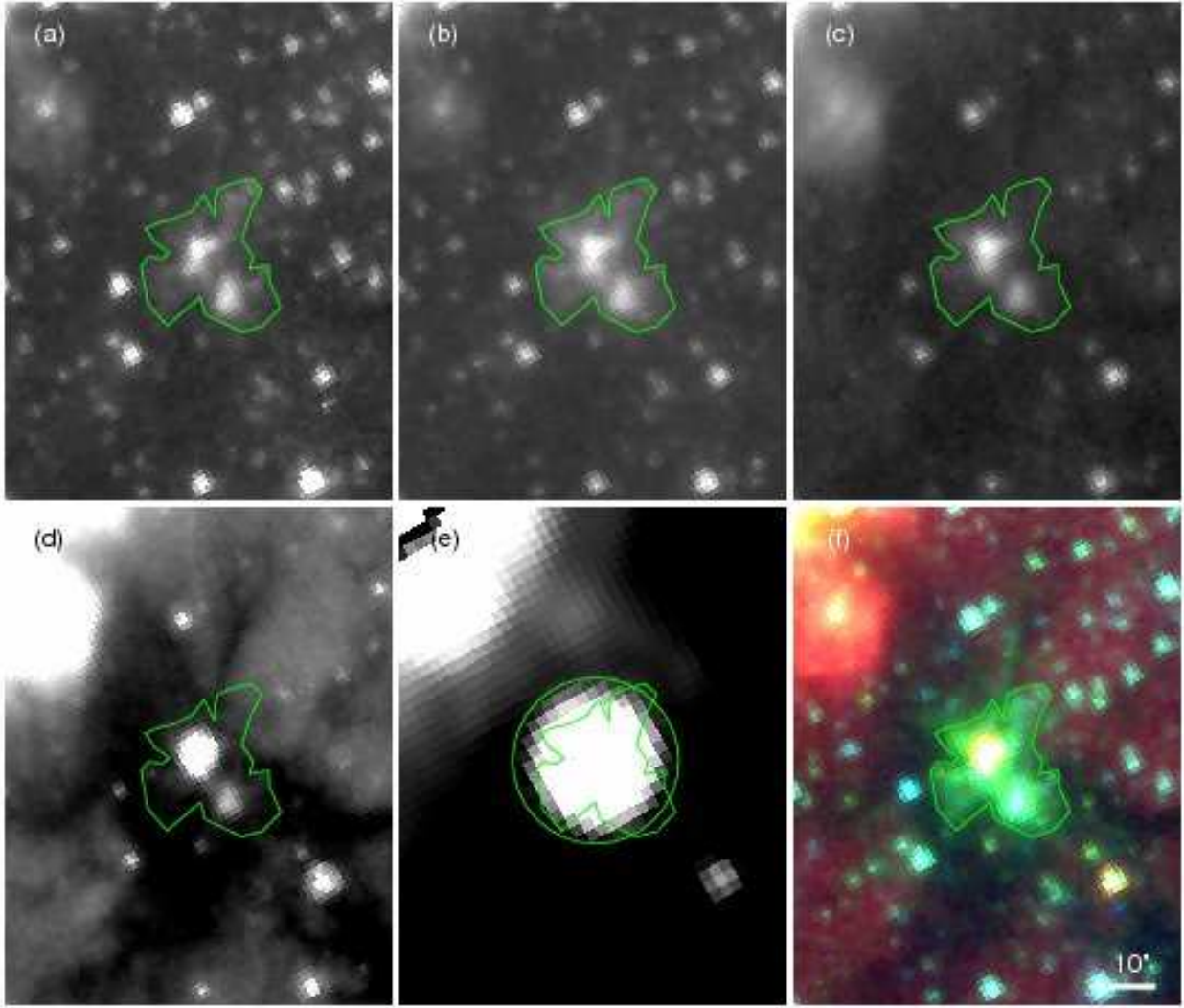}
\caption{(a-e): IRAC (a) 3.6 \um, (b) 4.5 \um, (c) 5.8 \um, (d) 8.0
\um\/ and (e) MIPS 24 \um\/ images of the EGO G11.92-0.61.  (f):
Three-color GLIMPSE IRAC image of the EGO G11.92-0.61 showing 8.0 \um
(red), 4.5~\um (green), and 3.6 \um (blue).  The polygonal source
aperture used for IRAC photometry is overlaid in all panels.  The
circular region used for MIPS photometry is overlaid on the 24 \um\/
image.  All panels are on the same spatial scale.  This is an example
of a source assigned to Table~\ref{fluxtable1}.}
\label{table1_example}
\end{figure}

\begin{figure}
\plotone{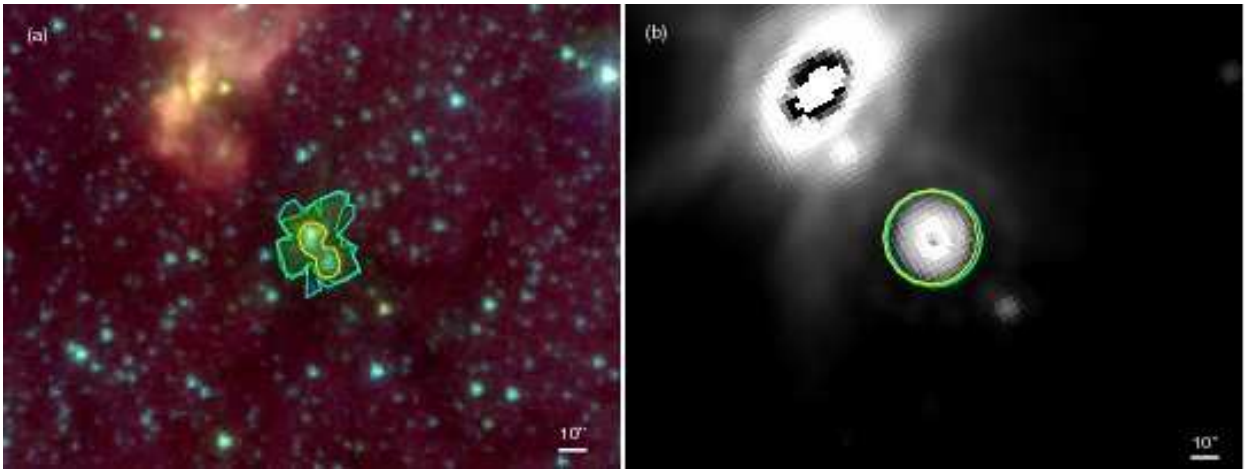}
\caption{(a) Three-color GLIMPSE IRAC image of the EGO G11.92-0.61 showing 8.0 \um
(red), 4.5~\um (green), and 3.6 \um (blue), with the 3 polygonal apertures
chosen for IRAC photometry by three different observers overlaid. (b) MIPS 24
\um\/ image of the EGO G11.92-0.61 with the 3 circular apertures chosen for
MIPS photometry by three different observers overlaid.  The MIPS counterpart
is slightly saturated, indicated by the dip in flux at its center.}
\label{3obs_apertures}
\end{figure}

\begin{figure} 
\plotone{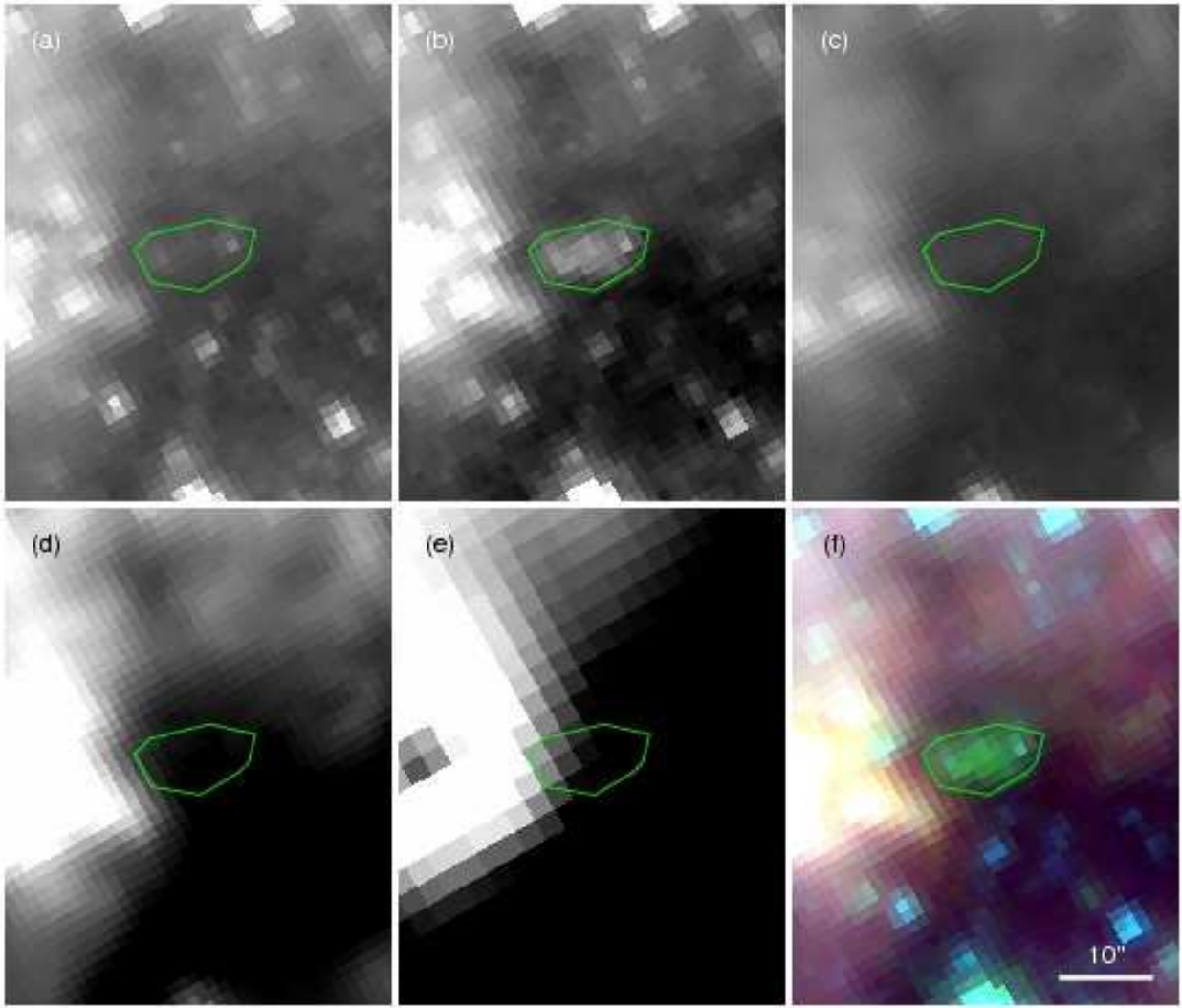}
\caption{Same as Figure~\ref{table1_example}, but for the EGO
  G10.29-0.13.  The polygonal source aperture used for IRAC photometry
  is overlaid in all panels.  This is an example of a source assigned to
  Table~\ref{fluxtable2} because of the lack of clear counterparts in
  bands other than [4.5].}
\label{table2_example1}
\end{figure}

\begin{figure} 
\plotone{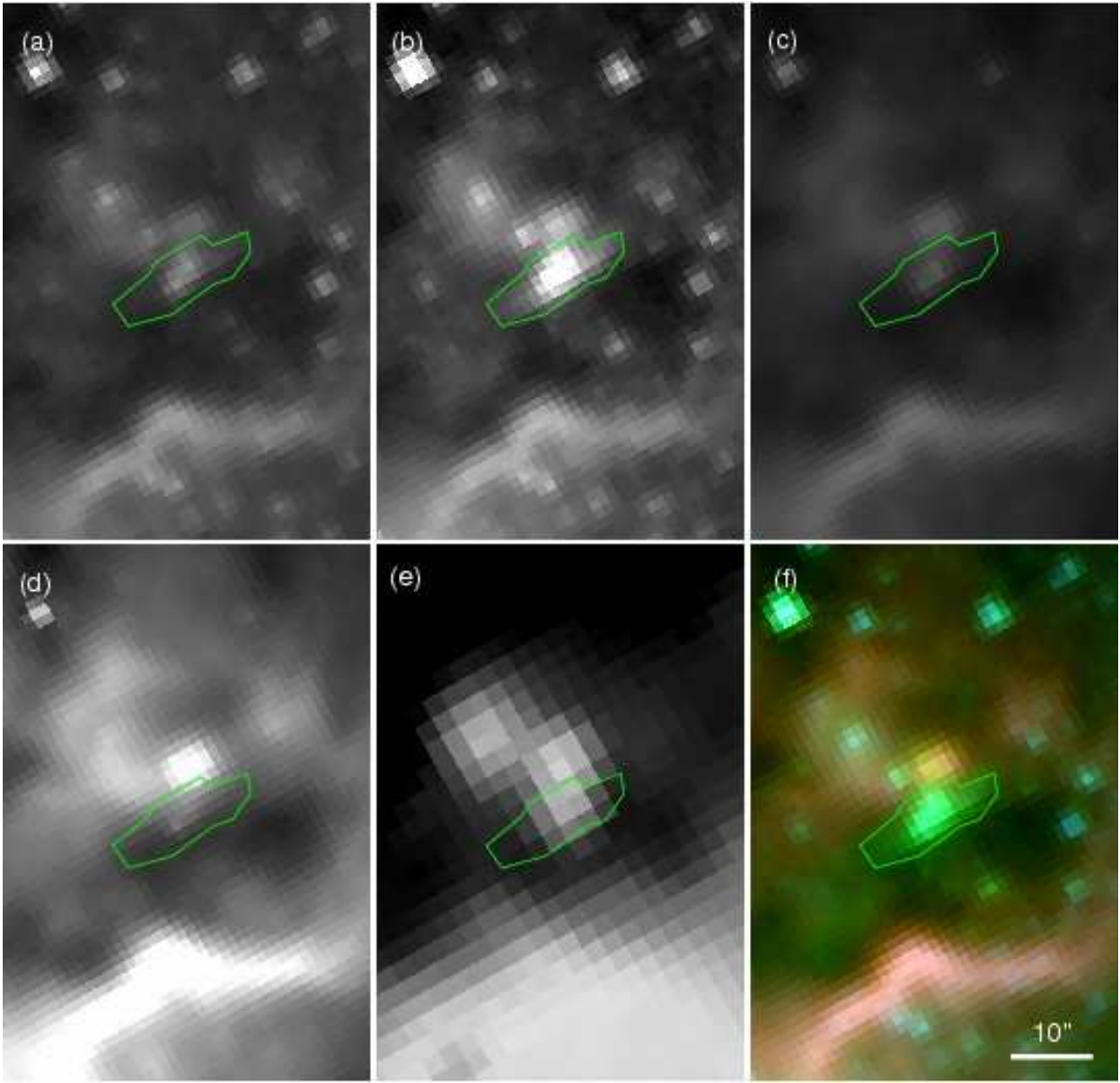}
\caption{Same as Figure~\ref{table1_example}, but for the EGO G10.34-0.14, an
  example of a source assigned to Table~\ref{fluxtable2} because of the lack
  of a clear counterpart at 8.0 \um\/ and confusion in the 24 \um\/
  image.  The polygonal source aperture used for IRAC photometry
  is overlaid in all panels.}
\label{table2_example2}
\end{figure}

\begin{figure} 
\plotone{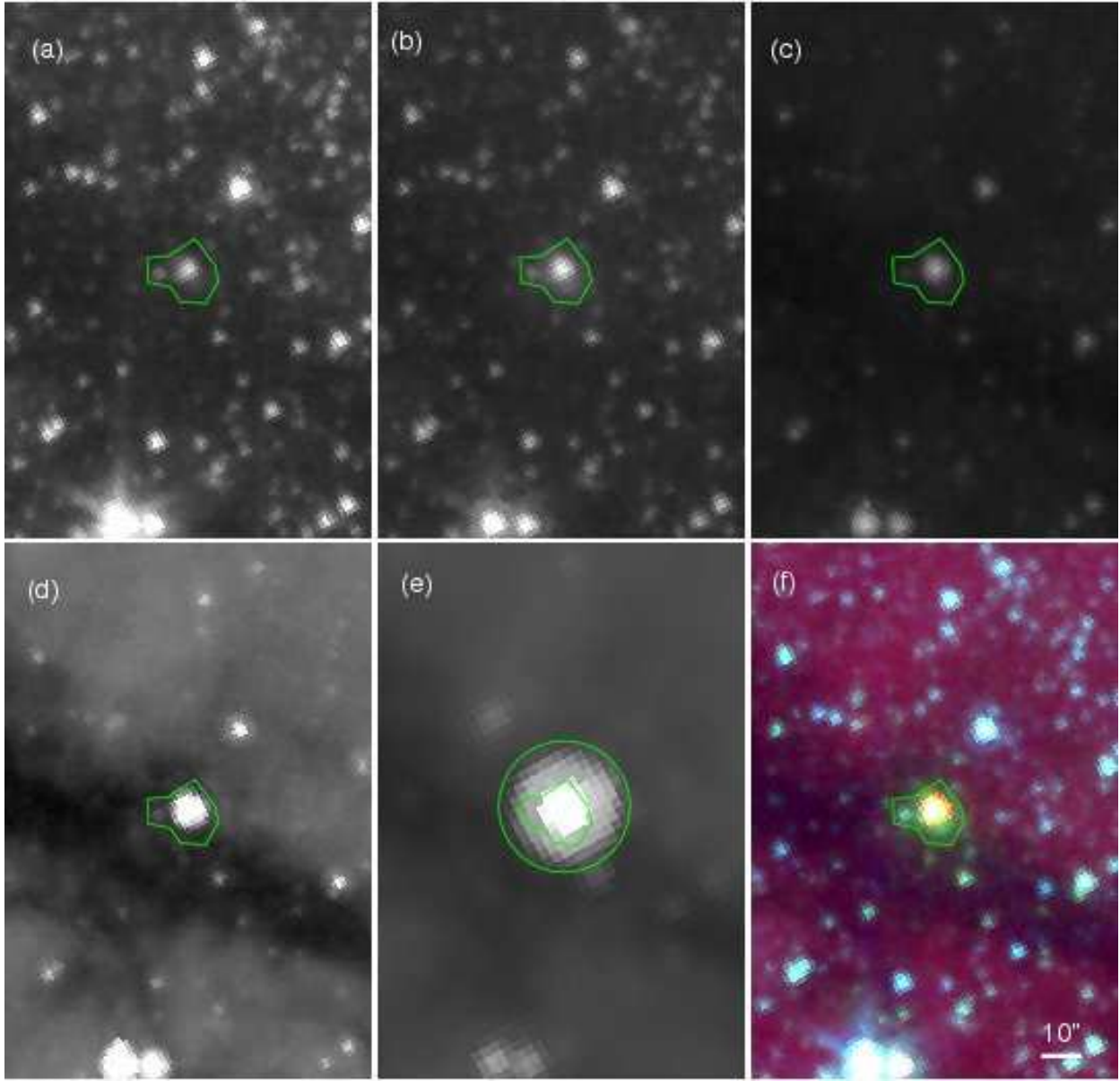}
\caption{Same as Figure~\ref{table1_example}, but for the EGO G11.11-0.11, an
example of a source assigned to Table~\ref{fluxtable1a} because, while clear
counterparts are visible in all bands, the green emission is not very extended
and surrounds a bright multiband IRAC source.   The polygonal source
aperture used for IRAC photometry is overlaid in all panels.  The
circular region used for MIPS photometry is overlaid on the 24 \um\/
image.}
\label{table1a_examples}
\end{figure}

\begin{figure} 
\plotone{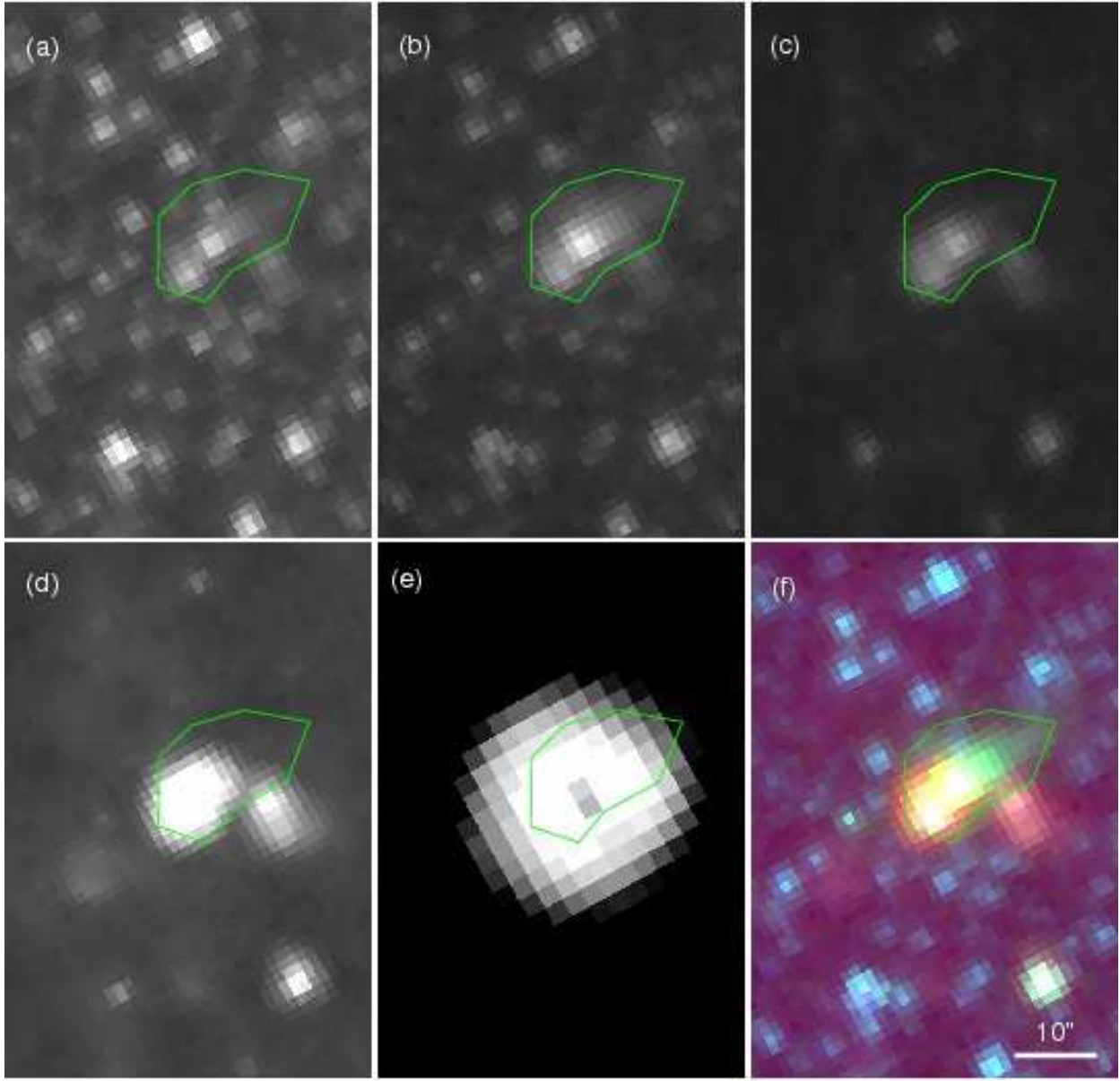}
\caption{Same as Figure~\ref{table1_example}, but for the EGO G12.20-0.03, an
example of a source assigned to Table~\ref{fluxtable3} because of the limited
extent of the 4.5 \um\/ emission and the degree of confusion in the region.  
At least 2 multiband IRAC sources are apparent, either
or both of which could be associated with the extended 4.5 \um\/ emission; 24
\um\/ emission from several IRAC sources appears to be blended in a single,
saturated, MIPS counterpart.   The polygonal source aperture used for IRAC photometry
  is overlaid in all panels.}
\label{table3_examples}
\end{figure}

Even at the resolution of IRAC, the star-forming regions in which some
EGOs are located are confused, leading to uncertainty in identifying
counterparts in all five mid-infrared (MIR) bands (3.6, 4.5, 5.8, 8.0,
and 24 \um) and in placing photometric apertures.  Since the spatial
resolution of MIPS at 24 \um\/ (\q6\arcsec) is about three times
poorer than that of IRAC, sources that are distinct in IRAC images may
be blended in MIPS 24 \um\/ images.  For each EGO, IRAC 3.6, 4.5, 5.8,
and 8.0 \um\/ images and the MIPS 24 \um\/ image were displayed with
the regions chosen for IRAC and MIPS photometry overlaid, as shown in
Figure~\ref{table1_example}a-e.  Equivalent figures for all catalogued
EGOs are available online.  If counterparts to the EGO are
distinguishable in all bands (as in Figure~\ref{table1_example}), then
five flux densities are reported (e.g. Tables~\ref{fluxtable1} and
\ref{fluxtable1a}).  Examples of EGOs assigned to
Tables~\ref{fluxtable1}-\ref{fluxtable3} are shown in, respectively,
Figure~\ref{table1_example} (Table~\ref{fluxtable1}),
Figures~\ref{table2_example1}-\ref{table2_example2}
(Table~\ref{fluxtable2}), Figure~\ref{table1a_examples}
(Table~\ref{fluxtable1a}), and Figure~\ref{table3_examples}
(Table~\ref{fluxtable3}).

\section{Discussion: The Nature of EGOs}

\subsection{Infrared Associations}

\begin{figure}
\plotone{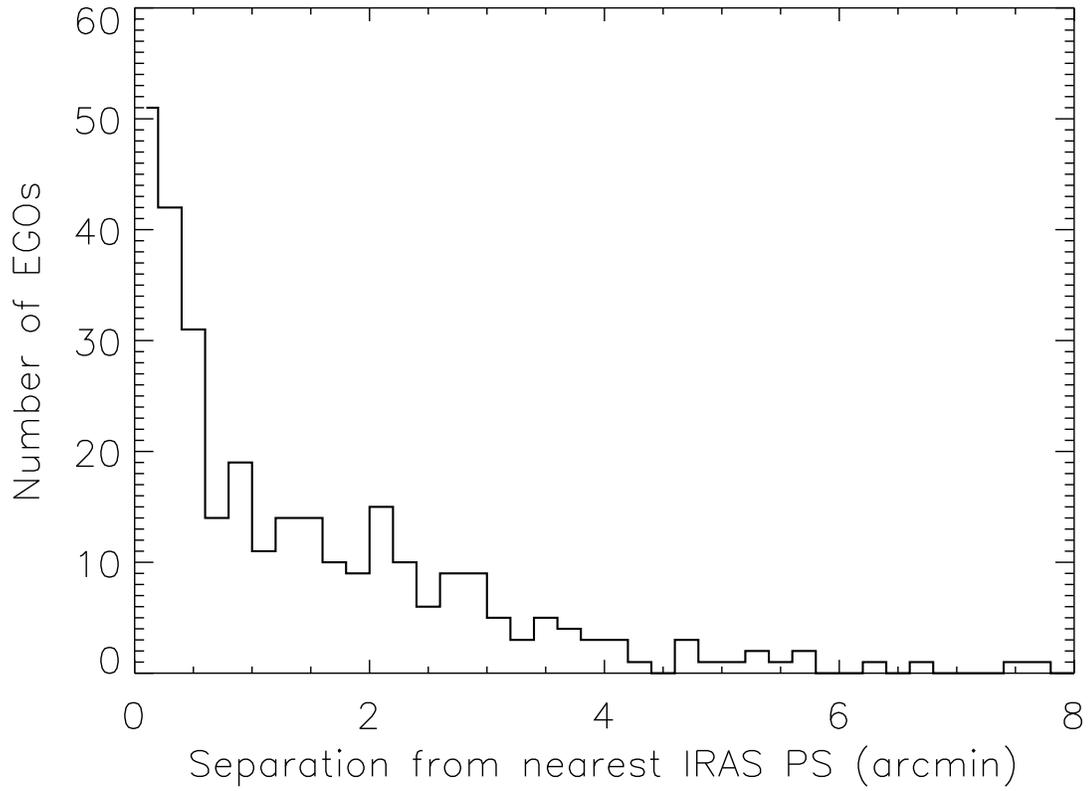}
\caption{Histogram, binned to 0.2\arcmin, of number of EGOs as a function of angular
  separation from the nearest \emph{IRAS} point source.  About \q 50\%
  of the catalogued EGO population is $>$1\arcmin\/ from the nearest
  \emph{IRAS} point source.}
\label{iras_histogram}
\end{figure}

\begin{figure}
\plotone{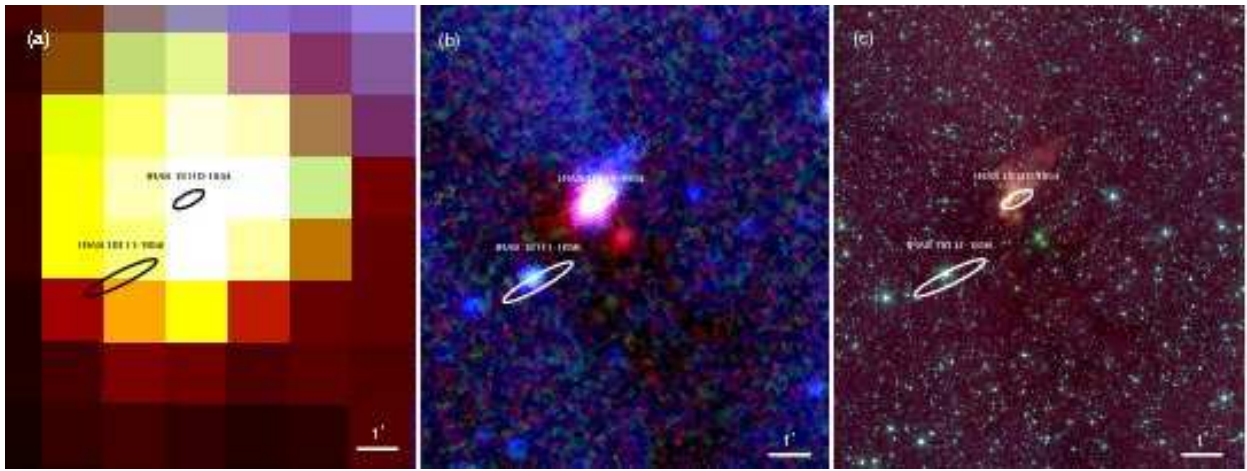}
\caption{The EGO G11.92-0.61 as seen by (a) \emph{IRAS} (red: 100 \um, green:
25 \um, blue: 12 \um), (b) \emph{MSX} (red: 21.4 \um, green: 12.1 \um, blue:
8.3 \um, and (c) \emph{Spitzer} IRAC (red: 8.0 \um, green: 4.5 \um, blue: 3.6
\um).  Error ellipses for the two IRAS point sources in the field are
overplotted.  All panels are on the same spatial scale.}
\label{3panel}
\end{figure}

The only data available for many of the EGOs in
Tables~\ref{fluxtable1}-\ref{fluxoutflowonly} come from IR all-sky
surveys, with \emph{Spitzer} and with previous satellites.  To assess
whether EGOs constitute a previously unrecognized population of MYSO
candidates, Table~\ref{irdc_table} lists the name and angular
separation of the nearest IRAS point source (catalog version 2.1) for
each catalogued EGO.  Figure~\ref{iras_histogram} graphically presents
the data in Table~\ref{irdc_table}.  While some EGOs may be identified
with IRAS point sources, and a few are well-known MYSOs (G12.91-0.26
(W33a), for example), the mean angular separation
between a catalogued EGO and the nearest IRAS point source is \q85\pp.
The mean angular separation is \q75\pp\/, \q84\pp\/, \q116\pp\/,
\q73\pp\/, and \q40\pp\/ for EGOs in Tables~\ref{fluxtable1},
~\ref{fluxtable2}, ~\ref{fluxtable1a}, ~\ref{fluxtable3} and
~\ref{fluxoutflowonly}, respectively.  Figure~\ref{3panel} shows the
EGO G11.92-0.61 (shown in
Figures~\ref{table1_example}-\ref{3obs_apertures}) as seen by
\emph{IRAS}, \emph{MSX}, and IRAC.  As illustrated by
Figure~\ref{3panel}, the fact that most EGOs are not \emph{IRAS} point
sources does not indicate a lack of emission at the wavelengths of the
\emph{IRAS} bands, but rather in many cases is attributable to
confusion and/or positional uncertainty at the resolution of
\emph{IRAS}.

Infrared dark clouds \citep[IRDCs,c.f.][]{Simon06a,Simon06b} have been shown
in recent years to be sites of the earliest stages of high mass star formation
\citep[e.g.][and references therein]{Rathborne07,Rathborne06}.  IRDCs are
readily visible against the diffuse background emission in GLIMPSE images,
particularly at 8.0 \um\/.  Many of these are also visible in \emph{MSX} 8.3
\um\/ (Band A) images and are catalogued in \citet{Simon06a} and
\citet{Simon06b}.  Small and/or filamentary IRDCs, however, may not be evident
at the 20\arcsec\/ resolution of \emph{MSX} Band A (\q 10 times poorer than
that of \emph{Spitzer} at 8 \um\/).  To take advantage of the higher
resolution afforded by \emph{Spitzer}, we have analyzed the correlation of
EGOs with IRDCs based on visual examination of 8.0 \um\/ GLIMPSE images.

\begin{figure}
\plotone{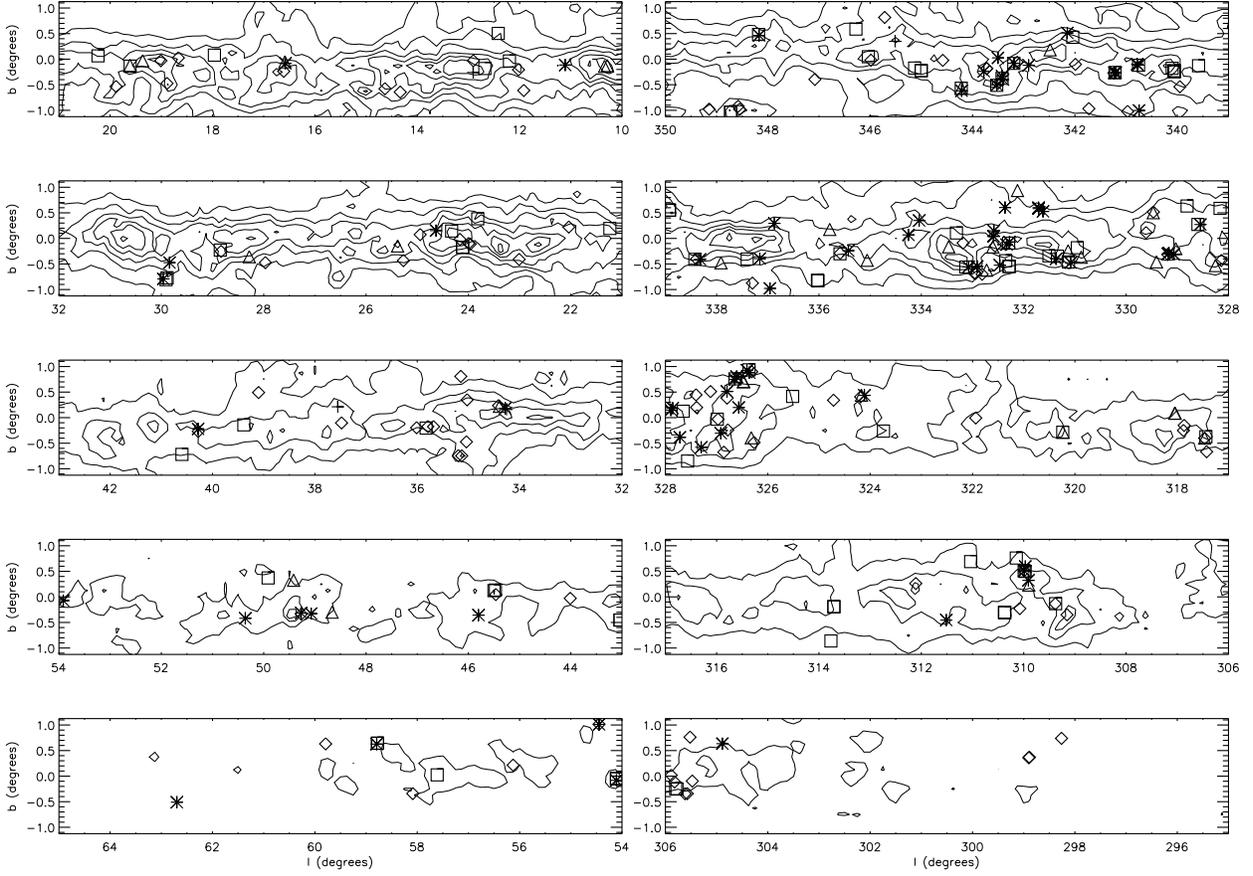}
\caption{Locations of EGOs overplotted on contours of integrated CO emission
from \citet{Dame01}.  EGOs in Table~\ref{fluxtable1} are shown as diamonds,
Table~\ref{fluxtable2} as squares, Table~\ref{fluxtable1a} as asterisks, Table
~\ref{fluxtable3} as triangles, and Table~\ref{fluxoutflowonly} as crosses.
CO contour levels are 39.5, 79.0,118.6,158.1,197.6,237.2 K \kms\/.  The
overall distribution pattern of EGOs with respect to CO is the same for
sources from all subcatalogs: the vast majority of EGOs coincide with CO
clouds.}
\label{coplot}
\end{figure}

The environment of each EGO included in
Tables~\ref{fluxtable1}-\ref{fluxoutflowonly} was examined in 8.0 \um\/
GLIMPSE images and IRDCs identified by visual examination; the results are
tabulated in Table~\ref{irdc_table}.  (Note that the 3-color IRAC images of
each EGO available online are not, in general, scaled to highlight the
presence or morphology of dark clouds.)  Overall, 67\% of the EGOs in
Tables~\ref{fluxtable1}-\ref{fluxoutflowonly} are in IRDCs, many of which are
small or filamentary.  Of our ``likely'' MYSO outflow candidate EGOs, 71\% are
associated with IRDCs: 72\% in Table ~\ref{fluxtable1} and 69\% in
Table~\ref{fluxtable2}.  Of our ``possible'' MYSO outflow candidate EGOs, 64\%
are associated with IRDCs: 77\% in Table ~\ref{fluxtable1a} and 54\% in
Table~\ref{fluxtable3}.  IRDCs are only visible in areas of high diffuse
galactic background emission, which generally decreases away from the Galactic
Plane.  Figure~\ref{coplot} shows the locations of EGOs overplotted on
integrated CO emission from \citet{Dame01}, with EGOs from each of
Tables~\ref{fluxtable1}-\ref{fluxoutflowonly} represented by a different
symbol.  The overall distribution pattern of EGOs with respect to CO is the
same for sources from all subcatalogs: the vast majority of EGOs coincide with
CO clouds.  For a subset of EGOs for which we had ancillary data, we
attempted to determine \vlsr\/ from the $^{13}$CO Galactic Ring Survey
\citep[GRS:][]{Jackson06}.  Due to the presence of multiple $^{13}$CO
components along the line of sight and the small angular size of EGOs,
however, the success rate was only \q50\%, which was not sufficiently
successful to attempt for the whole sample.  

The association of extended 4.5 \um\/ emission with known MYSOs (\S1), along
with the strong correlation of EGOs with IRDCs, suggests that extended 4.5
\um\/ emission in GLIMPSE images may be a good tracer of outflows specifically
from massive YSOs.  Extended 4.5 \um\/ emission can also originate from low
mass outflows \citep[e.g HH46/47,][]{NC04,Vel07}.  GLIMPSE is, however, a shallow
survey: the faintest EGOs identified in GLIMPSE and catalogued in
Tables~\ref{fluxtable1}-\ref{fluxoutflowonly} have surface brightnesses in 4.5
\um\/ images of $\gtrsim$ 4 MJy sr$^{-1}$.  In comparison, the bright knots in the
HH46/47 outflow have 4.5 \um\/ surface brightness of $\lesssim$ 2 MJy
sr$^{-1}$, and the majority of the diffuse green emission in NGC1333
\citep{Gutermuth08} has 4.5 \um\/ surface brightness of $\lesssim$ 4 MJy
sr$^{-1}$, with a few very bright areas of $\lesssim$ 10 MJy
sr$^{-1}$.  In both regions, the morphology of the diffuse 4.5 \um\/ emission is
mirrored in the other IRAC bands \citep[e.g. Fig.1 of][]{Vel07}, in contrast
to most of the EGOs in our sample (see \S4.3).  Relatively faint diffuse
4.5 \um\/ emission such as that seen in low-mass star-forming regions is
unlikely to have been identified in our search of the GLIMPSE images.

\subsection{Association with 6.7 GHz \meth\/ masers}

\begin{figure}
\plotone{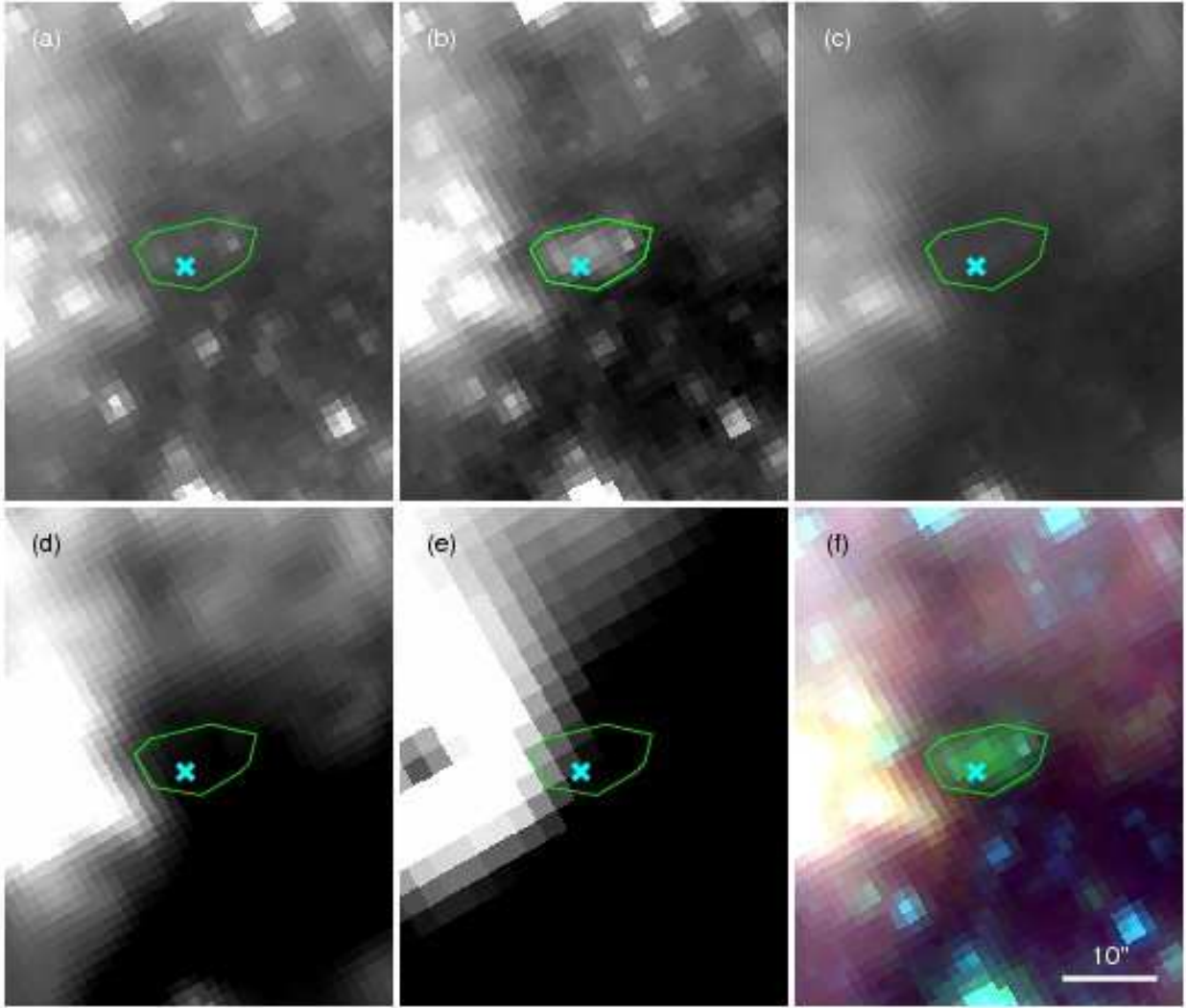}
\caption{Same as Figure~\ref{table2_example1}, but with a 6.7 GHz \meth\/ maser from
\citet{Walsh98} marked.  The region used for IRAC photometry is also
overplotted.  Similar images are available online for all EGOs
associated with 6.7 GHz \meth\/ masers in the surveys of
\citet{Caswell96},\citet{Walsh98} and \citet{Ellingsen06}.}
\label{maser_example}
\end{figure}

\begin{figure}
\plotone{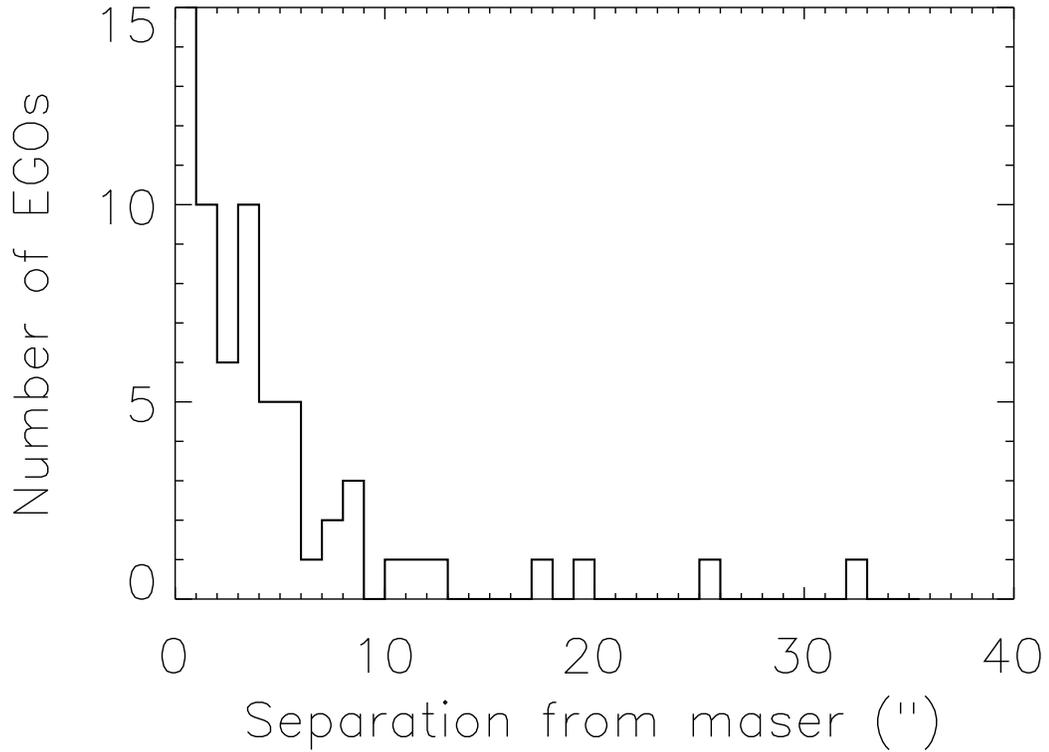}
\caption{Histogram, binned to 1.0\pp, of number of EGOs as a function of angular
  separation from associated 6.7 GHz \meth\/ masers.  This nominal
  offset is small ($\lesssim$5\pp\/ for the majority of EGOs).  This is
  consistent with our criteria for including an EGO in
  Table~\ref{maser_table}: a maser spatially coincident either
  with extended 4.5 \um\/ emission or with a 24 \um\/ or multiband
  IRAC source likely to be associated with extended 4.5 \um\/ emission.}
\label{maser_histogram}
\end{figure}

Further evidence that GLIMPSE-identified EGOs are massive comes from
the association of EGOs with 6.7 GHz Class II \meth\/ masers.  Class
II 6.7~GHz \meth\/ masers are indicative of massive star formation
\citep{Sz05, Ellingsen06}, as they are \emph{not} observed towards
low-mass YSOs \citep{Minier03}.  These masers are radiatively pumped
by IR photons emitted by warm dust in massive star-forming regions
\citep[e.g.][and references therein]{2005MNRAS.360..533C}, and are
thought to trace both the inner parts of outflows and disks
\citep[c.f.][]{2003MNRAS.341..277D}. We have compared our EGO catalog
with 6.7 GHz Class II \meth\/ maser catalogs of \citet{Ellingsen06},
\citet{Walsh98}, and \citet{Caswell96}.  To meaningfully assess the
correlation of EGOs and 6.7 GHz \meth\/ masers requires (a) maser
positions known to \q1\pp\/ or better, and (b) surveys covering
well-defined areas on the sky.  To our knowledge, the three surveys
cited \citep[and included in the analysis of GLIMPSE point sources
associated with 6.7 GHz \meth\/ masers conducted by][]{Ellingsen06}
are the only studies meeting these criteria that overlap the GLIMPSE
coverage.  Of these studies, those of \citet{Ellingsen06} and
\citet{Caswell96} cover (overlapping) ranges in galactic longitude.
That of \citet{Walsh98} targeted \emph{IRAS} sources; some EGOs
serendipitously fall within the (large) primary beam.
Table~\ref{maser_table} tabulates every EGO that falls within the
nominal coverage of one of these three surveys (25 EGOs from
Table~\ref{fluxtable1}, 23 from Table~\ref{fluxtable2}, 26 from
Table~\ref{fluxtable1a}, 35 from Table~\ref{fluxtable3}, and 2 from
Table~\ref{fluxoutflowonly}).  Column (1) lists the EGO name (galactic
coordinates), (2) the maser survey(s) that cover the EGO position, (3)
the \emph{IRAS} point source targeted by \citet{Walsh98} (if
relevant), (4) whether a maser was detected, (5) the nominal offset of
the maser position from the EGO position (from
Tables~\ref{fluxtable1}-\ref{fluxoutflowonly}), and (6) the distance.
The association of 6.7 GHz \meth\/ maser(s) with an EGO was evaluated
by plotting maser positions on images in each IRAC and the MIPS 24
\um\/ band and on three-color IRAC images of EGOs, as illustrated in
Figure~\ref{maser_example}a-f.  Only masers spatially coincident with
either (a) extended 4.5 \um\/ emission, or (b) a MIPS 24 \um\/ source
or IRAC multiband source likely to be associated with extended 4.5
\um\/ emission, are tabulated in Table~\ref{maser_table}.  In the
latter case, a ``?''  in column 4 indicates a possible association.
Figure~\ref{maser_histogram} graphically presents the nominal offsets
between EGO positions (as listed in
Tables~\ref{fluxtable1}-\ref{fluxoutflowonly}) and the maser positions
from the literature.  As shown in Figure~\ref{maser_histogram},
for the vast majority of EGOs this offset is small ($\lesssim$5\pp).  We
emphasize that EGOs are extended on the sky, and a maser with
a nominal positional offset of a few arcseconds generally lies within
the area of extended 4.5 \um\/ emission.  Figures equivalent to
Figure~\ref{maser_example}a-f are available online for all the sources
listed in Table~\ref{maser_table} to illustrate the range of 
projected locations of 6.7 GHz \meth\/ masers relative to extended
4.5 \um\/ emission.  Among the EGOs in
Table~\ref{maser_table} are examples of masers coincident with
multiband ``central'' sources, with 4.5 \um\/ emission without clear
counterparts in other MIR bands (e.g. the EGO shown in
Figures~\ref{table2_example1} and \ref{maser_example}), and with
bright MIPS 24 \um\/ point sources slightly offset from extended 4.5
\um\/ emission (e.g. G344.23-0.57).

The non-detections in Table~\ref{maser_table} should be treated with caution:
all of the maser surveys cited are flux-limited, and the sensitivity of the
\citet{Walsh98} and \citet{Caswell96} surveys is not uniform, but depends on the position
of a source within the interferometric primary beam.  This is reflected in the
citations for maser detections in column 4 of Table~\ref{maser_table}; in
several cases in which an EGO fell within the nominal coverage of multiple
maser surveys, a maser was detected in one survey but not in another.  The
fraction of EGOs within the maser survey areas found to be associated with 6.7 GHz
\meth\/ masers should thus be regarded as a lower limit: 73\% of 
``likely'' MYSO outflow candidate EGOs (61\% in
Table~\ref{fluxtable1} and 86\% in Table~\ref{fluxtable2}) and 27\% of
``possible'' MYSO outflow candidate EGOs (4\% in Table~\ref{fluxtable1a} and 44\% in
Table~\ref{fluxtable3}).  Both of the EGOs in Table~\ref{fluxoutflowonly} that
fall within the surveys' coverage have associated 6.7 GHz \meth\/ masers.

\citet{Ellingsen06} reports kinematic distances for detected 6.7 GHz \meth\/
masers; the near kinematic distances for masers associated with EGOs from
\citet{Ellingsen06} are listed in column 5 of Table~\ref{maser_table}.  The association of
EGOs with IRDCs supports the adoption of the near kinematic distance.  Where
two distances are listed, two masers in the \citet{Ellingsen06} sample appear
to be associated with the EGO.  The range of distances--from 2.6 to 5.3
kpc--is typical of distances to MSFR in the Galactic Plane and to IRDCs as
determined by \citet{Simon06b} from $^{13}$CO emission.

\subsection{MIR Colors}

\begin{figure}
\plotone{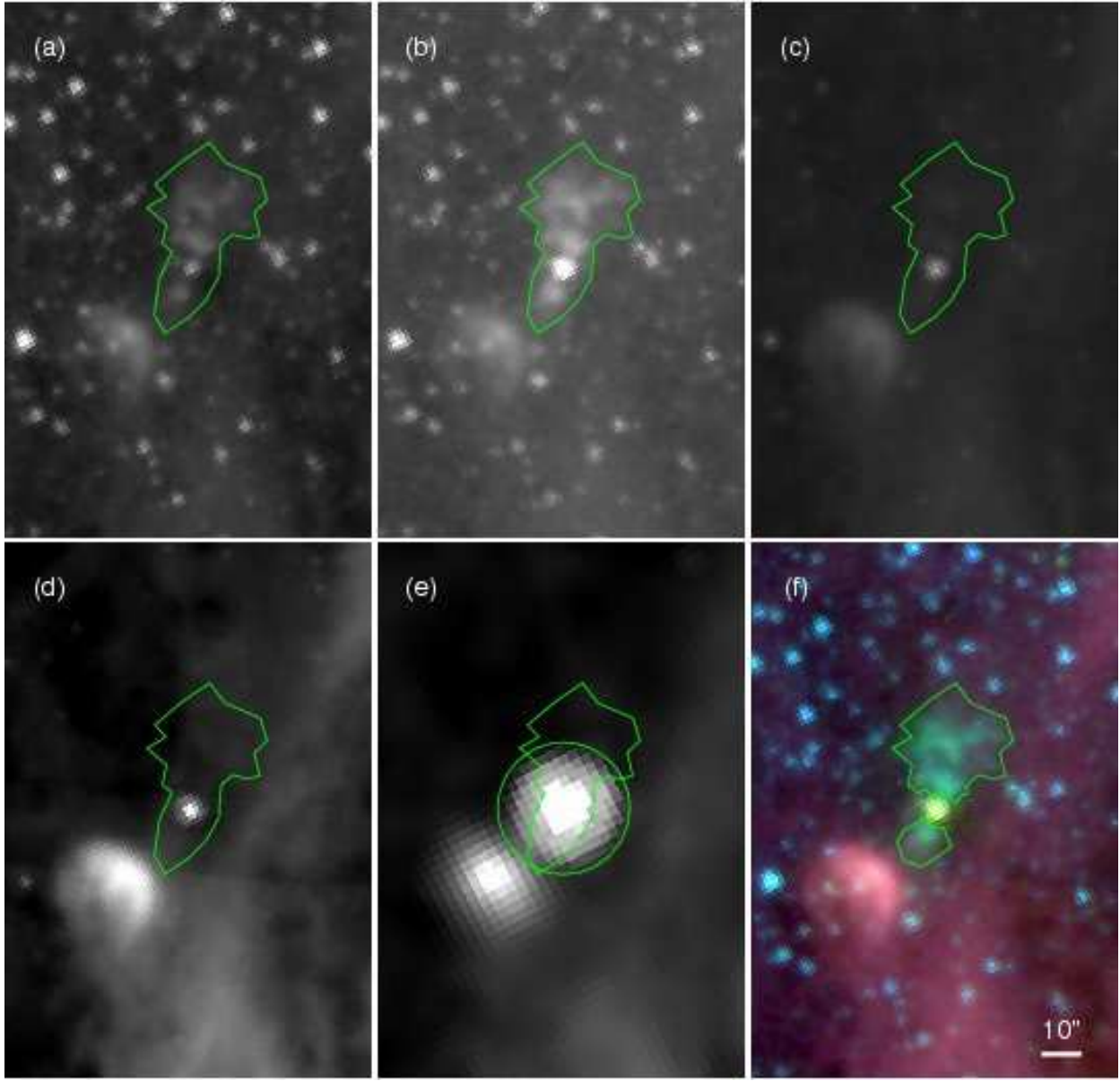}
\caption{(a-e): IRAC (a) 3.6 \um, (b) 4.5 \um, (c) 5.8 \um, (d) 8.0 \um\/ and
(e) MIPS 24 \um\/ images of the EGO G19.01-0.03.  The polygonal aperture used
for the photometry reported in Table~\ref{fluxtable1} is shown in all panels;
the aperture used for MIPS 24 \um\/ photometry is also shown in panel (e).
(f): Three-color GLIMPSE IRAC image of the EGO G19.01-0.03 showing 8.0 \um
(red), 4.5~\um (green), and 3.6 \um (blue) with the polygonal source apertures
used for ``outflow-only'' IRAC photometry reported in
Table~\ref{fluxoutflowonly} overlaid.  All panels are on the same spatial scale.  The
central point source dominates the flux density in the IRAC Table~\ref{fluxtable1}
aperture at wavelengths longer than 4.5 \um\/.}
\label{g19.01_sixpanel}
\end{figure}

\begin{figure}
\plotone{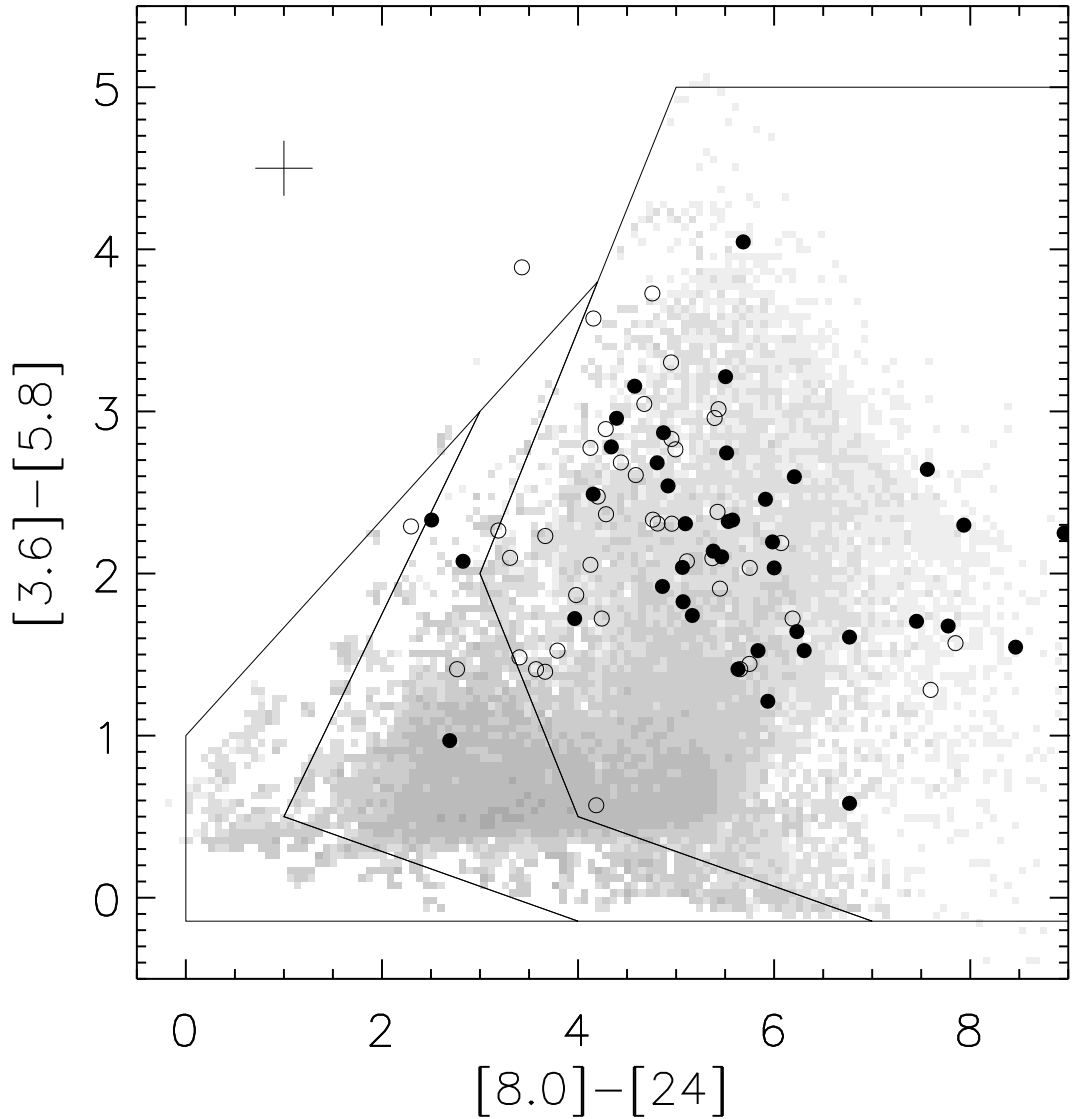}
\caption{Color-color plot of EGOs in Table~\ref{fluxtable1} (filled circles)
and Table~\ref{fluxtable1a} (open circles).  The error bar at top left was
derived from the average standard deviation of the measurements
(Tables~\ref{fluxtable1} and ~\ref{fluxtable1a}), converted to magnitudes and
propagated to colors.  The EGO points are overlaid on a grid of YSO models in
greyscale \citep{Robitaille06}.  The black lines indicate regions occupied
predominantly by models of various evolutionary stages: at right are the
youngest sources surrounded by infalling envelopes (Stage I); in the middle
are more evolved sources surrounded by disks (Stage II), and at left are
sources with low-mass disks (Stage III). Note that most EGOS fall in the
region occupied by the youngest protostar models.  EGOs with fluxes flagged as
saturated or as upper limits in Tables~\ref{fluxtable1} and ~\ref{fluxtable1a}
were excluded from the plot.}
\label{colorcolor}
\end{figure}

As illustrated in Figure~\ref{g19.01_sixpanel}, the integrated flux
densities reported in Tables~\ref{fluxtable1} and \ref{fluxtable1a}
are likely dominated by emission from the central source of an EGO, at
least at wavelengths $>$4.5 \um.  As a result, the integrated flux
densities may be used, with caution, to infer the properties of the
sources responsible for driving the hypothesized outflows traced by
extended 4.5 \um\/ emission.  Figure~\ref{colorcolor} shows that EGOs
are found in the region of color-color space occupied by very young
YSOs, surrounded by substantial accreting envelopes
\citep{Robitaille06}.  Detailed spectral energy distribution (SED)
fitting has not been attempted because without long-wavelength
information the SEDs are poorly constrained, and the resolution of
MIPSGAL at 70 \um\/ is too poor to resolve EGOs in crowded regions.

The EGO shown in Figure~\ref{g19.01_sixpanel} is a rare example of an
EGO in which a central source is sufficiently distinct that extended
``outflow'' emission may be photometered separately.  As shown in
Figure~\ref{g19.01_sixpanel}a-e, the extended 4.5 \um\/ emission
evident in Figure~\ref{g19.01_sixpanel}b only has a comparably
extended counterpart in the 3.6 \um\/ band.
Table~\ref{fluxoutflowonly} reports 3.6 \um\/ and 4.5 \um\/ integrated
flux densities for the ``outflow-only'' regions shown in
Figure~\ref{g19.01_sixpanel}f and for four additional sources in which
``outflow-only'' regions can be distinguished (shown in online
Figure~\ref{g19.01_sixpanel}).  The [3.6]-[4.5] colors of the extended
emission in these five sources fall within the range determined for
``the shocked outflow nebulosity'' in the DR21 outflow by
\citet{Smith06} and redward in [3.6]-[4.5] space of the non-outflow
``surrounding nebulosity'' \citep[c.f. Figure 7 of][]{Smith06}.

\section{Conclusions}

We have catalogued over 300 EGOs, identified by their extended 4.5
\um\/ emission, in GLIMPSE images.  The suggestion that extended 4.5
\um\/ emission in GLIMPSE traces outflows specifically from
\emph{massive} YSOs is supported by the strong correlation of EGOs
with IRDCs and 6.7 GHz Class II \meth\/ masers, both associated with
early stages of massive star formation.  The mid-infrared colors of
EGOs lie in regions of color-color space occupied by young protostars
still embedded in infalling envelopes.  Taken together, this evidence
suggests that \emph{Spitzer} images (e.g. the GLIMPSE I, II and 3D
surveys and other Galactic programs) may be used to identify a large
number of new MYSO candidates \emph{that are actively accreting}, a
crucial category of sources for constraining models of high-mass star
formation.

The catalog presented here represents a source list for further study.  Most
of the candidates have not been previously studied, and broadband
\emph{Spitzer} images are insufficient to ensure a homogeneous pool of
candidates.  We have selected a sample of \q30 northern EGOs for extensive
followup observations designed to test the hypothesis that 4.5 \um\/ emission
traces shocked gas in outflows from massive YSOs.  This sample was chosen to
span a range of 4.5 \um\/ morphologies and associations (or lack thereof) with
other tracers of massive star formation.  We will report the results of
these studies in subsequent papers.

\acknowledgments

This work is based on observations made with the {\it Spitzer Space
Telescope}, which is operated by the Jet Propulsion Laboratory, California
Institute of Technology under a contract with NASA.  This work was based on
data taken by the GLIMPSE and MIPSGAL \emph{Spitzer Space Telescope} Legacy
Programs.  This research has made use of NASA's Astrophysics Data System
Bibliographic Services and the SIMBAD database operated at CDS, Strasbourg,
France.  Support for this work was provided by NASA/JPL through contract
1289524.  C.J.C. is supported by a National Science Foundation Graduate
Research Fellowship.

\clearpage
\begin{deluxetable}{lccccccccccc}
\rotate
\tablewidth{0pt}
\tablecaption{MIR Integrated Flux Densities:``likely'' MYSO outflow candidates\label{fluxtable1}} 
\tablehead{ 
\colhead{Name} & 
\multicolumn{2}{c}{J2000 Coordinates} &
\colhead{area((\arcsec)$^{2}$)} &
\multicolumn{4}{c}{Integrated flux(mJy)} &
\colhead{area((\arcsec)$^{2}$)} &
\colhead{Integrated flux(mJy)} &
\colhead{Saturated?} &
\colhead{Upper limit?} \\
\colhead{} &
\colhead{$\alpha$ ($^{\rm h}~~^{\rm m}~~^{\rm s}$)} &
\colhead{$\delta$ ($^{\circ}~~{\arcmin}~~{\arcsec}$)} &
\colhead{IRAC} &
\colhead{[3.6]} &
\colhead{[4.5]} &
\colhead{[5.8]} &
\colhead{[8.0]} &
\colhead{MIPS} &
\colhead{[24]} &
\colhead{} &
\colhead{}
}
\tablecolumns{12}
\tabletypesize{\scriptsize}
\setlength{\tabcolsep}{0.05in}
\startdata
G11.92-0.61&18 13 58.1&-18 54 17&583&90.7(1.7)&334.4(3.6)&355.8(23.7)&197.8(36.6)&962&7331.9 (154.6)&Y, 24& \\ 
G12.02-0.21&18 12 40.4&-18 37 11&73&3.9(2.1)&14.1(1.3)&16.6(11.1)&11.3(1.7)&132&113.5 (26.1)&N& \\ 
G12.91-0.03&18 13 48.2&-17 45 39&75&4.9(0.1)&42.7(0.7)&36.8(8.4)&24.3(5.4)&213&2168.7 (49.6)&N&8,24 \\ 
G14.33-0.64&18 18 54.4&-16 47 46&163&12.1(0.8)&53.8(0.7)&123.7(8.3)&186.3(22.8)&346&4918.9 (73.7)&Y, 24&8,24 \\ 
G14.63-0.58&18 19 15.4&-16 30 07&79&6.3(0.2)&24.2(0.3)&44.3(1.0)&50.9(5.5)&772&1563.4 (105.0)&N&24 \\ 
G16.61-0.24&18 21 52.7&-14 35 51&170&15.0(0.3)&43.6(2.8)&43.7(6.4)&21.2(13.3)&559&333.3 (34.7)&N& \\ 
G18.67+0.03&18 24 53.7&-12 39 20&111&17.8(2.2)&56.1(2.1)&84.7(0.5)&72.0(13.8)&662&3921.3 (60.5)&N&24 \\ 
G18.89-0.47&18 27 07.9&-12 41 36&85&6.9(2.2)&25.4(1.8)&24.3(21.5)&38.6(20.9)&138&475.8 (74.1)&N& \\ 
G19.01-0.03&18 25 44.8&-12 22 46&721&88.7(15.2)&307.2(12.6)&471.5(80.9)&570.1(88.8)&835&3468.4 (287.3)&N& \\ 
G19.88-0.53&18 29 14.7&-11 50 23&423&145.6(24.6)&566.2(15.9)&910.6(78.8)&790.8(135.6)&783&5055.1 (14.8)&N& \\ 
G22.04+0.22&18 30 34.7&-09 34 47&75&1.7(1.4)&14.0(0.2)&19.3(3.5)&8.8(6.4)&536&3293.5 (193.8)&N&24 \\ 
G23.01-0.41&18 34 40.2&-09 00 38&272&33.1(2.3)&225.5(3.6)&446.7(32.9)&221.5(30.7)&973&6189.0 (481.4)&Y,24& \\ 
G23.96-0.11&18 35 22.3&-08 01 28&184&30.7(1.6)&172.7(3.0)&245.2(12.4)&92.2(3.0)&852&1636.5 (99.1)&N& \\ 
G24.00-0.10&18 35 23.5&-07 59 32&37&0.7(0.4)&8.9(0.7)&16.7(1.1)&13.7(3.4)&69&294.0 (1.6)&N& \\ 
G24.17-0.02&18 35 25.0&-07 48 15&46&1.9(0.6)&8.9(0.8)&6.9(2.8)&5.8(5.8)&115&114.4 (4.1)&N& \\ 
G24.94+0.07&18 36 31.5&-07 04 16&223&22.3(3.7)&55.3(1.6)&68.1(1.7)&49.8(12.1)&818&1383.1 (116.5)&N& \\ 
G25.27-0.43&18 38 57.0&-07 00 48&262&14.6(3.2)&43.3(2.4)&32.9(21.7)&16.2(4.4)&156&190.8 (4.2)&N& \\ 
G27.97-0.47&18 44 03.6&-04 38 02&202&8.6(1.2)&48.0(0.3)&83.1(12.1)&102.6(30.8)&484&1820.7 (36.4)&N&24 \\ 
G28.83-0.25&18 44 51.3&-03 45 48&575&52.3(15.2)&171.0(19.5)&115.4(52.9)&142.5(19.5)&478&3635.9 (64.6)&N&8,24 \\ 
G34.41+0.24&18 53 17.9&+01 25 25&114&7.2(0.4)&21.4(0.4)&24.9(4.3)&16.5(10.7)&864&6227.1 (32.4)&Y,24& \\ 
G35.03+0.35&18 54 00.5&+02 01 18&845&161.6(11.8)&501.1(7.3)&511.9(217.8)&279.6(438.5)&1089&10181.9 (231.7)&Y,24&3,4,5,8,24 \\ 
G35.04-0.47&18 56 58.1&+01 39 37&154&8.5(1.7)&25.4(0.8)&17.6(0.5)&11.8(5.8)&426&188.8 (16.0)&N&3,4,5,8,24 \\ 
G35.13-0.74&18 58 06.4&+01 37 01&145&97.5(0.8)&181.1(3.0)&327.2(8.0)&452.2(43.7)&104&1305.5 (31.1)&N&8,24 \\ 
G35.15+0.80&18 52 36.6&+02 20 26&351&97.6(2.0)&133.7(2.3)&275.6(1.4)&367.9(9.4)&714&1880.2 (33.8)&N&3,4,5,8,24 \\ 
G35.20-0.74&18 58 12.9&+01 40 33&1066&490.3(9.3)&1080.3(2.7)&1623.1(38.2)&2879.6(97.3)&9971&31704.5 (530.1)&Y,24& \\ 
G35.68-0.18&18 57 05.0&+02 22 00&117&5.2(0.4)&23.3(0.8)&8.7(0.2)&7.9(2.5)&449&485.6 (40.5)&N&8,24 \\ 
G35.79-0.17&18 57 16.7&+02 27 56&48&3.2(0.6)&12.2(0.3)&14.5(1.3)&11.4(6.5)&668&1303.1 (39.0)&N& \\ 
G36.01-0.20&18 57 45.9&+02 39 05&109&5.4(1.6)&18.0(1.7)&12.1(0.8)&5.8(1.5)&109&59.3 (3.5)&N& \\ 
G37.48-0.10&19 00 07.0&+03 59 53&107&7.5(1.2)&19.0(0.8)&40.9(4.0)&60.7(7.1)&714&1096.3 (364.0)&N& \\ 
G39.10+0.49&19 00 58.1&+05 42 44&120&11.9(0.3)&29.2(1.0)&23.0(0.9)&9.7(1.8)&876&1439.8 (9.1)&N& \\ 
G40.28-0.27&19 05 51.5&+06 24 39&114&17.2(0.2)&73.4(0.4)&127.3(3.4)&107.7(8.1)&2604&3221.2 (87.5)&N&24 \\ 
G44.01-0.03&19 11 57.2 &+09 50 05&117&30.0(0.7)&31.7(0.6)&29.6(1.1)&11.8(5.8)&63&16.0 (3.8)&N& \\ 
G45.47+0.05&19 14 25.6&+11 09 28&248&65.0(3.7)&183.3(2.8)&143.7(20.5)&91.5(69.5)&847&15702.4 (806.6)&Y,24& \\ 
G49.27-0.34&19 23 06.7&+14 20 13&441&82.5(11.2)&666.3(8.0)&1737.2(42.9)&1880.6(144.9)&1319&8677.1 (174.5)&Y,24& \\ 
G54.45+1.02&19 28 25.7&+19 32 20&50&4.5(0.2)&9.4(0.0)&15.6(1.3)&22.7(1.8)&52&93.2 (2.4)&N&24 \\ 
G56.13+0.22&19 34 51.5&+20 37 28&62&3.6(0.1)&9.0(0.1)&17.0(0.4)&22.3(0.6)&207&153.5 (2.4)&N&24 \\ 
G58.09-0.34&19 41 03.9&+22 03 39&79&10.0(0.1)&12.2(0.3)&7.2(0.3)&1.1(0.8)&582&57.4 (1.4)&N& \\ 
G59.79+0.63&19 41 03.1&+24 01 15&121&7.8(0.9)&26.9(0.7)&45.8(3.3)&41.9(10.7)&518&417.4 (29.0)&N& \\ 
G298.26+0.74&12 11 47.7&-61 46 21&356&317.1(1.6)&605.6(0.3)&648.8(4.5)&536.3(8.0)&876&18226.2 (51.3)&Y,24&24 \\ 
G298.89+0.37&12 16 37.9&-62 13 41&35&1.9(0.1)&3.4(0.2)&2.7(0.4)&0.6(0.5)&86&20.2 (3.0)&N& \\ 
G298.90+0.36&12 16 43.2&-62 14 25&278&71.7(1.6)&125.7(0.7)&160.1(4.6)&162.0(7.9)&2540&3293.1 (211.9)&N&24 \\ 
G305.48-0.10&13 13 45.8&-62 51 28&376&107.8(4.3)&275.4(0.9)&358.5(16.5)&304.1(35.9)&639&5052.2 (568.7)&Y, 24& \\ 
G305.52+0.76&13 13 29.3&-61 59 53&228&20.7(1.0)&48.7(0.1)&55.8(3.1)&22.8(3.8)&783&645.6 (45.5)&N& \\ 
G305.57-0.34&13 14 49.1&-63 05 38&141&20.3(0.7)&35.9(0.6)&25.0(7.1)&13.3(2.9)&536&343.9 (15.0)&N& \\ 
G305.62-0.34&13 15 11.5&-63 05 30&85&15.4(0.8)&27.1(0.3)&27.2(3.9)&19.0(8.7)&478&1082.8 (25.1)&N& \\ 
G305.82-0.11&13 16 48.6&-62 50 35&84&15.6(0.2)&57.5(0.4)&64.6(1.3)&50.0(9.2)&484&1062.2 (185.6)&N&24 \\ 
G305.89+0.02&13 17 15.5&-62 42 24&124&36.7(0.9)&120.4(0.2)&166.4(14.1)&158.5(19.6)&622&3639.0 (77.5)&N&24 \\ 
G309.15-0.35&13 45 51.3&-62 33 46&60&1.8(0.1)&7.2(0.2)&6.6(2.9)&8.0(4.9)&81&9.3 (3.3)&N& \\ 
G309.38-0.13(a)&13 47 23.9&-62 18 12&131&8.2(0.3)&20.3(1.2)&26.2(4.4)&9.4(9.6)&645&3852.7 (54.8)&N& \\ 
G309.99+0.51(a)&13 51 12.2&-61 32 09&95&3.9(1.0)&8.8(0.1)&7.8(2.6)&15.7(6.4)&92&69.3 (1.1)&N& \\ 
G310.08-0.23&13 53 23.0&-62 14 13&128&53.1(1.2)&121.6(0.8)&183.6(1.0)&161.7(9.9)&605&2959.2 (95.5)&N& \\ 
G312.11+0.26&14 08 49.3&-61 13 25&98&10.2(0.6)&29.0(0.4)&17.3(8.7)&10.2(0.7)&708&2713.6 (44.8)&N& \\ 
G317.42-0.67&14 51 59.2&-60 06 06&102&22.5(1.0)&106.3(0.5)&222.7(3.8)&242.0(8.7)&668&1173.9 (8.3)&N&24 \\ 
G317.46-0.40b&14 51 19.6&-59 50 51&89&10.8(1.6)&55.1(0.4)&103.4(2.4)&117.2(14.9)&893&3933.1 (118.5)&Y,24&24 \\ 
G317.87-0.15&14 53 16.3&-59 26 36&72&6.4(0.1)&13.1(0.2)&10.5(2.4)&8.0(5.4)&161&192.8 (13.7)&N& \\ 
G317.88-0.25&14 53 43.5&-59 31 35&73&11.4(0.7)&22.1(0.1)&19.4(4.5)&19.0(1.1)&46&17.5 (1.0)&N&24 \\ 
G321.94-0.01&15 19 43.3&-57 18 06&102&18.9(1.1)&98.3(0.6)&190.8(2.6)&173.5(5.9)&570&4416.1 (57.4)&N&24 \\ 
G324.19+0.41&15 31 38.0&-55 42 36&115&25.9(0.6)&32.2(0.7)&53.0(3.6)&53.7(7.2)&138&207.0 (1.0)&N&8,24? \\ 
G324.72+0.34&15 34 57.5&-55 27 26&462&26.4(3.6)&86.5(2.5)&65.6(14.9)&12.8(7.1)&708&896.7 (10.8)&N&24 \\ 
G326.27-0.49&15 47 10.8&-55 11 12&547&147.4(11.8)&494.5(7.1)&921.6(14.1)&784.5(48.6)&737&3139.6 (173.1)&N&8,24 \\ 
G326.31+0.90&15 41 35.9&-54 03 42&53&13.5(0.5)&27.2(0.2)&36.0(4.3)&45.3(3.7)&121&68.2 (27.5)&N& \\ 
G326.78-0.24&15 48 55.2&-54 40 37&295&207.5(0.7)&614.1(0.5)&1085.6(8.5)&1415.1(18.1)&634&9509.5 (101.5)&Y, 24& \\ 
G326.79+0.38&15 46 20.9&-54 10 45&406&171.0(2.9)&305.0(3.2)&348.5(25.0)&289.1(23.0)&697&3762.9 (26.1)&N& \\ 
G326.86-0.67&15 51 13.6&-54 58 03&226&13.0(0.9)&35.8(2.1)&62.8(4.9)&13.6(11.2)&109&130.9 (11.2)&N& \\ 
G326.97-0.03&15 49 03.2&-54 23 37&46&2.4(0.1)&11.1(0.5)&16.8(1.2)&11.2(3.1)&104&67.3 (7.9)&N&24 \\ 
G327.12+0.51&15 47 32.7&-53 52 39&1048&365.3(13.0)&964.5(4.3)&2544.5(11.3)&4347.8(47.4)&962&19185.4 (143.3)&Y, 24&5.8,8 \\ 
G327.39+0.20&15 50 18.5&-53 57 07&170&67.5(0.1)&196.5(0.4)&259.9(4.4)&242.4(24.9)&547&3322.0 (23.5)&N&24 \\ 
G327.40+0.44&15 49 19.3&-53 45 10&318&235.6(1.7)&876.7(5.3)&1964.7(7.4)&2829.9(10.5)&841&11960.0 (150.3)&Y,24&8,24 \\ 
G328.14-0.43&15 56 57.6&-53 57 48&89&64.9(1.3)&146.9(0.7)&127.8(4.4)&23.8(7.8)&564&2567.3 (15.0)&N& \\ 
G329.18-0.31&16 01 47.4&-53 11 44&189&18.2(1.6)&90.0(1.4)&89.7(4.4)&31.7(11.7)&662&3969.5 (28.7)&N&24 \\ 
G329.47+0.52&15 59 36.6&-52 22 55&117&10.3(0.8)&28.3(1.1)&33.7(2.6)&6.5(3.8)&536&1167.9 (30.8)&N& \\ 
G329.61+0.11&16 02 03.1&-52 35 33&274&110.8(0.9)&498.7(0.7)&880.5(21.7)&1431.7(60.6)&3312&8092.8 (438.8)&N&24 \\ 
G332.35-0.12&16 16 07.0&-50 54 30&308&198.6(3.4)&429.4(1.9)&636.1(26.7)&612.0(70.5)&605&5557.2 (162.1)&Y,24&24 \\ 
G332.56-0.15&16 17 12.1&-50 47 14&92&20.7(1.6)&51.0(2.2)&34.6(2.5)&21.1(26.5)&115&782.4 (197.7)&N& \\ 
G332.81-0.70&16 20 48.1&-51 00 15&160&63.2(4.6)&182.2(5.1)&178.6(14.6)&180.9(81.0)&570&3104.6 (458.5)&N& \\ 
G332.94-0.69&16 21 18.9&-50 54 10&240&44.7(3.8)&215.0(2.7)&336.7(36.1)&354.9(43.1)&559&2689.8 (550.2)&N&8 \\ 
G332.96-0.68&16 21 22.9&-50 52 58&164&34.7(0.6)&47.9(0.8)&283.0(6.0)&705.3(11.8)&444&3622.0 (56.8)&Y,24&8 \\ 
G333.18-0.09&16 19 45.6&-50 18 34&206&129.2(5.5)&338.1(4.0)&344.5(9.7)&107.0(85.4)&121&1270.5 (137.7)&N&8 \\ 
G335.59-0.29&16 30 58.5&-48 43 51&982&132.1(11.9)&483.8(9.9)&466.7(54.5)&111.3(134.4)&2160&10230.0 (244.0)&Y,24& \\ 
G337.30-0.87&16 40 31.3&-47 51 31&199&56.5(1.3)&111.2(0.7)&103.8(6.8)&38.5(22.3)&547&1356.2 (67.6)&N& \\ 
G338.39-0.40&16 42 41.2&-46 43 40&907&175.8(5.4)&336.0(12.4)&300.4(20.1)&117.6(63.5)&553&3298.4 (63.0)&N&8,24 \\ 
G339.95-0.54&16 49 07.9&-45 37 59&170&41.7(1.1)&140.3(0.8)&188.4(12.5)&231.6(6.9)&766&7877.0 (182.9)&N& \\ 
G340.97-1.02&16 54 57.3&-45 09 04&304&12.2(5.7)&95.5(0.1)&75.0(30.2)&44.4(17.8)&680&591.3 (89.5)&N&8 \\ 
G341.24-0.27&16 52 37.3&-44 28 09&153&20.8(1.2)&57.6(0.1)&60.6(13.9)&38.0(18.5)&536&1189.0 (64.0)&N&8,24 \\ 
G341.73-0.97&16 57 23.1&-44 31 35&334&73.3(2.8)&213.2(0.8)&295.7(6.2)&230.9(9.1)&530&1184.9 (13.1)&N& \\ 
G341.99-0.10&16 54 32.8&-43 46 45&276&48.8(2.7)&162.0(2.3)&192.8(7.6)&89.1(11.7)&691&2304.0 (30.4)&N& \\ 
G343.12-0.06&16 58 16.6&-42 52 04&894&196.2(2.6)&875.6(2.7)&1254.9(21.7)&796.6(90.2)&3519&19300.2 (512.4)&Y,24& \\ 
G343.50-0.47&17 01 18.4&-42 49 36&204&44.4(0.7)&204.9(1.0)&301.2(8.7)&216.7(8.4)&657&3482.4 (41.8)&N&24 \\ 
G343.72-0.18(a)&17 00 48.3&-42 28 25&197&33.6(6.6)&60.2(1.3)&55.3(3.5)&32.6(11.7)&495&905.6 (21.7)&N&24 \\ 
G343.72-0.18(b)&17 00 48.1&-42 28 37&59&4.5(0.6)&15.4(0.1)&18.3(4.4)&31.4(4.9)&472&1092.5 (107.0)&N&24 \\ 
G344.58-0.02&17 02 57.7&-41 41 54&259&58.0(1.1)&205.8(2.1)&320.0(5.2)&430.1(11.7)&3663&12302.6 (43.5)&Y, 24&8,24 \\ 
G345.72+0.82&17 03 06.4&-40 17 09&1231&149.2(6.6)&694.6(2.6)&1400.5(29.6)&1371.7(80.6)&3698&7499.2 (73.8)&Y, 24&8,24 \\ 
G345.99-0.02&17 07 27.6&-40 34 45&76&4.7(0.2)&20.2(1.5)&35.4(1.6)&11.8(1.2)&207&514.3 (8.1)&N&24 \\ 
G347.08-0.40&17 12 27.2&-39 55 18&72&51.2(0.3)&100.7(0.0)&101.8(2.2)&51.8(6.0)&173&2286.4 (5.5)&N&24 \\ 
G348.55-0.98&17 19 20.9&-39 03 55&438&198.9(3.4)&442.9(11.4)&650.0(39.7)&680.9(162.8)&564&6324.5 (109.0)&N&8,24 \\ 
G348.58-0.92&17 19 10.7&-39 00 23&301&75.9(1.7)&216.8(2.8)&246.0(2.9)&154.2(21.2)&576&5098.8 (118.3)&Y, 24& \\ 
G349.15-0.98&17 21 04.8&-38 34 25&279&177.9(2.6)&355.6(2.1)&627.0(3.1)&918.5(6.6)&3197&4861.4 (159.7)&Y, 24&24 \\ 
\enddata
\end{deluxetable}

\begin{deluxetable}{lcccc}
\tablewidth{0pt}
\tablecaption{[4.5] Integrated Flux Densities: "likely" MYSO outflow candidates \label{fluxtable2}} 
\tablehead{ 
\colhead{Name} & 
\multicolumn{2}{c}{J2000 Coordinates} &
\colhead{area((\arcsec)$^{2}$)} &
\colhead{Integrated flux(mJy)} \\
\colhead{} &
\colhead{$\alpha$ ($^{\rm h}~~^{\rm m}~~^{\rm s}$)} &
\colhead{$\delta$ ($^{\circ}~~{\arcmin}~~{\arcsec}$)} &
\colhead{IRAC} &
\colhead{[4.5]} 
}
\tablecolumns{5}
\tabletypesize{\scriptsize}
\setlength{\tabcolsep}{0.05in}
\startdata
G10.29-0.13&18 08 49.3&-20 05 57.3&69&13.1(1.4) \\ 
G10.34-0.143&18 09 00.0&-20 03 35&94&72.7(4.3) \\ 
G16.59-0.05&18 21 09.1&-14 31 48&128&50.3(0.7) \\ 
G19.36-0.03&18 26 25.8&-12 03 56.9&153&115.0(2.1) \\ 
G19.61-0.12&18 27 13.6&-11 53 20&102&32.4(5.1) \\ 
G25.38-0.15&18 38 08.1&-06 46 53&46&46.4(0.9) \\ 
G28.28-0.36&18 44 13.2&-04 18 04&112&47.5(3.3) \\ 
G34.39+0.22&18 53 19.0&+01 24 08&307&20.4(4.3) \\ 
G48.66-0.30&19 21 48.0&+13 49 21&120&2.0(0.9) \\ 
G49.42+0.33&19 20 59.1&+14 46 53&98&12.7(0.9) \\ 
G309.90+0.23&13 51 00.4&-61 49 53&109&16.0(0.2) \\ 
G317.46-0.40a&14 51 19.6&-59 50 41&39&15.3(0.4) \\ 
G318.04+0.09&14 53 39.2&-59 09 10&84&19.6(0.2) \\ 
G318.05+0.09&14 53 42.6&-59 08 49&202&419.5(1.3) \\ 
G320.23-0.28&15 09 52.6&-58 25 36&170&82.4(0.8) \\ 
G324.17+0.44&15 31 24.6&-55 41 30&84&16.4(0.1) \\ 
G326.32-0.39&15 47 04.8&-55 04 51&91&19.1(0.0) \\ 
G326.47+0.70&15 43 15.4&-54 07 14&69&7.8(0.2) \\ 
G326.48+0.70&15 43 17.5&-54 07 11&164&92.5(0.3) \\ 
G326.61+0.80(b)&15 43 34.6&-53 58 00&84&17.7(0.6) \\ 
G328.25-0.53&15 57 59.7&-53 58 00&523&866.3(3.0) \\ 
G329.03-0.20&16 00 30.6&-53 12 34&397&81.6(1.7) \\ 
G329.41-0.46&16 03 32.4&-53 09 26&219&258.9(3.1) \\ 
G329.47+0.50&15 59 41.0&-52 23 28&219&59.0(1.3) \\ 
G330.88-0.37&16 10 19.9&-52 06 13&88&41.9(1.4) \\ 
G331.13-0.24&16 10 59.8&-51 50 19&46&21.5(0.9) \\ 
G332.12+0.94&16 10 30.4&-50 18 05&121&43.7(0.7) \\ 
G332.60-0.17&16 17 29.4&-50 46 13&82&14.2(2.0) \\ 
G332.73-0.62&16 20 02.8&-51 00 32&186&80.1(3.2) \\ 
G333.47-0.16&16 21 20.2&-50 09 50&117&125.8(0.9) \\ 
G335.06-0.43&16 29 23.1&-49 12 28&157&46.5(1.9) \\ 
G335.79+0.18&16 29 47.1&-48 15 47&134&93.7(1.9) \\ 
G337.91-0.48&16 41 10.3&-47 08 06&613&1981.1(6.0) \\ 
G341.22-0.26(a)&16 52 32.2&-44 28 38&158&24.2(4.4) \\ 
G342.48+0.18&16 55 02.6&-43 13 01&179&42.6(2.3) \\ 
G344.23-0.57&17 04 07.1&-42 18 42&81&18.7(0.5) \\ 
\enddata
\end{deluxetable}

\begin{deluxetable}{lccccccccccc}
\rotate
\tablewidth{0pt}
\tablecaption{MIR Integrated Flux Densities: "possible" MYSO outflow candidates \label{fluxtable1a}} 
\tablehead{ 
\colhead{Name} & 
\multicolumn{2}{c}{J2000 Coordinates} &
\colhead{area((\arcsec)$^{2}$)} &
\multicolumn{4}{c}{Integrated flux(mJy)} &
\colhead{area((\arcsec)$^{2}$)} &
\colhead{Integrated flux(mJy)} &
\colhead{Saturated?} &
\colhead{Upper limit?} \\
\colhead{} &
\colhead{$\alpha$ ($^{\rm h}~~^{\rm m}~~^{\rm s}$)} &
\colhead{$\delta$ ($^{\circ}~~{\arcmin}~~{\arcsec}$)} &
\colhead{IRAC} &
\colhead{[3.6]} &
\colhead{[4.5]} &
\colhead{[5.8]} &
\colhead{[8.0]} &
\colhead{MIPS} &
\colhead{[24]} &
\colhead{} &
\colhead{}
}
\tablecolumns{12}
\tabletypesize{\scriptsize}
\setlength{\tabcolsep}{0.05in}
\startdata
G11.11-0.11&18 10 28.3&-19 22 31&177&22.2(3.7)&200.7(1.6)&386.7(6.1)&263.1(24.7)&812&1542.6 (80.2)&N& \\ 
G16.58-0.08&18 21 15.0&-14 33 02&82&63.7(0.5)&63.5(0.5)&122.6(6.5)&83.1(23.8)&559&497.9 (16.6)&N&24 \\ 
G24.63+0.15&18 35 40.1&-07 18 35&52&8.4(0.9)&59.8(1.7)&103.1(1.5)&57.9(3.5)&530&405.1 (125.9)&N&24 \\ 
G29.84-0.47&18 47 28.8&-02 58 03&78&1.4(0.8)&14.0(0.2)&11.2(2.2)&7.4(3.2)&202&36.4 (12.4)&N& \\ 
G29.96-0.79&18 48 50.0&-03 00 21&232&82.6(1.5)&128.7(1.5)&227.2(17.4)&367.2(2.4)&167&199.8 (19.9)&N&8 \\ 
G34.28+0.18&18 53 15.0&+01 17 11&157&13.3(3.2)&47.7(1.5)&68.1(7.3)&44.8(47.0)&444&499.7 (5.7)&N& \\ 
G40.28-0.22&19 05 41.3&+06 26 13&225&80.6(0.8)&236.3(0.3)&249.1(0.5)&204.8(1.9)&3214&7733.8 (198.0)&N&24 \\ 
G45.80-0.36&19 16 31.1&+11 16 11&65&19.1(0.2)&63.0(0.3)&103.8(1.4)&108.1(3.9)&109&854.1 (1.2)&N&24 \\ 
G49.07-0.33&19 22 41.9&+14 10 12&48&23.3(0.6)&85.9(1.0)&135.3(0.3)&66.1(18.9)&98&381.0 (57.8)&N& \\ 
G49.27-0.32&19 23 02.2&+14 20 52&30&2.9(0.7)&5.6(0.7)&3.9(3.5)&6.9(1.6)&524&855.0 (180.1)&N& \\ 
G50.36-0.42&19 25 32.8&+15 15 38&82&14.4(0.5)&20.5(0.4)&21.3(0.6)&11.6(2.3)&576&245.4 (0.3)&N& \\ 
G53.92-0.07&19 31 23.0&+18 33 00&105&58.9(0.2)&108.0(0.6)&149.9(0.8)&171.5(0.9)&616&1043.1 (45.1)&N&24 \\ 
G54.11-0.08&19 31 48.8&+18 42 57&58&54.2(0.7)&130.4(0.7)&244.4(5.2)&359.0(15.5)&559&2748.6 (542.0)&N& \\ 
G54.45+1.01&19 28 26.4&+19 32 15&91&23.7(0.3)&75.8(0.4)&147.9(1.8)&176.6(6.2)&588&1880.5 (26.4)&N&24 \\ 
G58.79+0.63&19 38 55.3&+23 09 04&94&7.0(0.1)&12.3(0.1)&13.8(0.5)&16.4(0.6)&230&88.6 (1.6)&N& \\ 
G62.70-0.51&19 51 51.1&+25 57 40&99&47.8(0.2)&64.7(0.2)&77.1(0.3)&74.4(1.2)&570&190.0 (3.8)&N& \\ 
G304.89+0.64&13 08 12.1&-62 10 22&72&45.7(0.5)&102.9(0.2)&165.8(2.7)&175.1(10.2)&3185&7645.7 (32.7)&Y, 24& \\ 
G309.91+0.32&13 50 53.9&-61 44 22&251&61.1(1.1)&129.4(2.1)&96.7(18.5)&20.5(3.2)&467&581.0 (192.8)&N&24 \\ 
G309.97+0.59&13 50 52.6&-61 27 46&56&2.2(0.5)&4.0(0.1)&3.2(0.7)&3.0(1.7)&81&9.1 (0.9)&N& \\ 
G309.99+0.51(b)&13 51 12.7&-61 32 22&49&3.1(0.4)&7.8(0.3)&10.7(2.0)&2.3(1.7)&63&33.2 (2.7)&N& \\ 
G311.51-0.45&14 05 46.1&-62 04 49&190&397.1(1.5)&776.0(0.8)&1183.6(12.8)&1507.7(14.4)&3070&10139.1 (1231.5)&Y,24&24 \\ 
G324.11+0.44&15 31 05.0&-55 43 39&354&58.9(1.3)&185.2(0.7)&207.4(4.8)&171.3(8.3)&737&1537.6 (7.1)&N& \\ 
G326.36+0.88&15 41 55.4&-54 02 55&50&2.8(0.8)&9.2(0.5)&12.2(3.2)&13.8(5.1)&104&74.7 (57.1)&N& \\ 
G326.41+0.93&15 41 59.4&-53 59 03&343&75.7(3.1)&323.6(3.1)&468.2(24.5)&333.4(19.0)&772&2551.9 (208.0)&N&8 \\ 
G326.57+0.20&15 45 53.4&-54 27 50&50&4.4(0.4)&17.9(0.1)&25.3(0.8)&16.8(1.2)&132&273.4 (2.5)&N& \\ 
G326.61+0.80(a)&15 43 35.1&-53 57 36&150&20.2(0.5)&72.9(0.8)&115.6(4.0)&181.7(19.0)&484&2248.1 (16.9)&N&24 \\ 
G326.61+0.80(c)&15 43 36.2&-53 57 51&125&92.7(1.4)&162.4(0.5)&174.2(6.1)&150.1(16.0)&501&1816.6 (18.0)&N&24 \\ 
G326.65+0.75&15 44 00.9&-53 58 45&140&40.5(0.7)&196.8(0.4)&300.3(4.5)&218.2(15.8)&507&2110.8 (27.9)&N&24 \\ 
G326.80+0.51&15 45 48.6&-54 04 30&81&3.7(0.3)&10.2(0.4)&5.8(3.8)&6.7(3.0)&52&10.0 (2.6)&N& \\ 
G326.92-0.31&15 49 56.2&-54 38 29&166&11.0(0.9)&60.6(0.1)&97.8(9.6)&84.3(30.2)&599&551.4 (113.4)&N&24 \\ 
G327.30-0.58&15 53 11.2&-54 36 48&89&24.5(1.5)&158.8(1.7)&341.8(8.8)&364.9(34.1)&369&2589.1 (86.5)&N&24 \\ 
G327.72-0.38&15 54 32.3&-54 11 55&37&5.1(0.2)&16.0(0.6)&37.6(3.3)&54.2(6.1)&75&113.0 (19.7)&N&8,24 \\ 
G327.86+0.19&15 52 49.2&-53 40 07&52&7.6(0.1)&15.9(0.4)&26.2(1.5)&28.8(3.6)&305&339.6 (45.1)&N&8,24 \\ 
G327.89+0.15&15 53 10.3&-53 40 28&89&10.9(0.5)&38.6(0.2)&57.9(4.0)&43.4(4.9)&179&215.0 (37.9)&N& \\ 
G328.55+0.27&15 56 01.5&-53 09 44&157&85.1(2.5)&268.8(1.2)&368.4(10.7)&318.4(31.4)&812&4483.8 (57.9)&Y,24& \\ 
G329.07-0.31(a)&16 01 11.7&-53 16 00&158&79.8(0.9)&277.9(1.7)&423.4(5.7)&424.4(16.2)&668&4192.3 (39.4)&N&24 \\ 
G329.07-0.31(b)&16 01 09.9&-53 16 02&217&57.5(0.6)&322.8(0.4)&795.3(5.5)&1054.1(8.7)&795&6590.2 (45.9)&Y,24&8,24 \\ 
G329.16-0.29&16 01 33.6&-53 11 15&36&4.3(0.3)&11.8(0.0)&16.0(1.1)&11.8(6.3)&75&106.5 (1.1)&N&24 \\ 
G329.18-0.30&16 01 44.9&-53 11 15&167&15.5(0.8)&66.0(1.3)&104.6(4.1)&50.3(19.3)&645&581.7 (14.6)&N&8 \\ 
G331.08-0.47&16 11 46.9&-52 02 31&111&6.6(1.0)&26.6(1.7)&60.2(5.2)&58.7(11.9)&415&629.1 (79.4)&N& \\ 
G331.37-0.40&16 12 48.1&-51 47 30&86&2.1(1.3)&9.7(0.8)&4.3(3.4)&5.0(1.4)&213&166.9 (11.3)&N& \\ 
G331.62+0.53&16 09 56.8&-50 56 25&363&47.6(5.6)&117.2(2.0)&233.2(15.1)&444.2(25.0)&714&2955.7 (75.8)&N& \\ 
G331.71+0.58&16 10 06.3&-50 50 29&161&10.7(1.0)&26.5(0.6)&16.6(5.0)&10.6(6.2)&334&245.5 (11.7)&N& \\ 
G331.71+0.60&16 10 01.9&-50 49 33&520&122.5(2.2)&315.3(1.4)&421.9(5.5)&395.8(22.8)&876&3720.8 (38.6)&N& \\ 
G332.28-0.07&16 15 35.1&-50 55 36&53&9.8(0.3)&29.2(1.9)&24.1(8.8)&14.9(6.1)&150&551.9 (71.1)&N&24 \\ 
G332.33-0.12&16 16 03.3&-50 55 34&48&8.1(0.9)&20.4(0.1)&26.9(2.5)&23.7(6.6)&98&68.3 (7.8)&N&8 \\ 
G332.36+0.60&16 13 02.4&-50 22 39&207&102.2(2.1)&237.5(1.2)&368.7(1.6)&382.8(8.4)&714&2216.9 (93.4)&N& \\ 
G332.47-0.52&16 18 26.5&-51 07 12&544&149.5(4.6)&420.9(2.8)&1029.5(73.0)&1898.5(105.5)&870&8610.4 (182.3)&Y,24& \\ 
G332.58+0.15&16 16 00.6&-50 33 30&69&7.1(0.7)&22.0(0.9)&24.3(4.0)&19.2(6.4)&167&203.9 (17.9)&N& \\ 
G332.59+0.04(a)&16 16 30.4&-50 37 41&30&3.4(0.4)&8.8(0.6)&4.8(2.8)&6.0(6.6)&52&22.4 (0.9)&N& \\ 
G332.59+0.04(b)&16 16 30.1&-50 37 50&52&7.9(0.6)&20.5(0.5)&27.2(7.2)&41.1(6.5)&52&38.1 (1.9)&N& \\ 
G332.91-0.55&16 20 32.6&-50 49 46&75&14.5(0.6)&47.4(0.8)&42.5(5.9)&8.4(8.3)&138&240.2 (70.8)&N& \\ 
G333.08-0.56&16 21 20.9&-50 43 05&151&37.7(3.6)&277.3(0.7)&558.8(37.7)&570.9(68.6)&513&1500.1 (187.1)&N& \\ 
G334.04+0.35&16 21 36.9&-49 23 28&177&31.0(1.5)&101.9(0.4)&171.8(1.4)&218.7(8.7)&645&2344.9 (18.3)&N& \\ 
G334.25+0.07&16 23 45.2&-49 26 32&86&8.8(0.3)&23.4(0.6)&24.1(2.8)&8.7(6.4)&138&201.0 (11.3)&N& \\ 
G335.43-0.24&16 30 05.8&-48 48 44&252&38.0(2.3)&298.1(3.5)&481.7(32.1)&266.4(28.8)&570&2380.3 (39.3)&N& \\ 
G336.87+0.29&16 33 40.3&-47 23 32&147&21.8(3.2)&66.6(2.5)&62.1(22.5)&39.2(6.2)&478&612.7 (15.1)&N& \\ 
G336.96-0.98&16 39 37.5&-48 10 58&125&27.1(0.2)&64.6(0.3)&64.1(7.2)&42.0(2.0)&628&708.6 (67.2)&N& \\ 
G337.16-0.39&16 37 49.6&-47 38 50&222&228.3(0.7)&536.9(1.8)&1136.5(15.3)&2237.2(43.8)&1140&5589.0 (144.0)&N&24 \\ 
G338.32-0.41&16 42 27.5&-46 46 57&193&39.5(1.7)&102.5(1.2)&101.5(6.2)&30.9(5.9)&743&1479.0 (46.8)&N&8,24 \\ 
G340.75-1.00&16 54 04.0&-45 18 50&320&101.3(6.3)&239.0(1.1)&336.4(28.2)&552.6(14.9)&3554&17376.0 (370.9)&Y,24&24 \\ 
G340.77-0.12&16 50 17.5&-44 43 54&111&13.1(0.6)&34.3(0.9)&36.3(3.9)&28.6(13.3)&472&358.6 (32.1)&N& \\ 
G340.78-0.10&16 50 14.7&-44 42 31&91&6.7(1.0)&22.0(0.6)&25.4(6.2)&20.0(11.3)&726&1737.9 (58.7)&N&24 \\ 
G341.20-0.26&16 52 27.8&-44 29 29&58&23.3(0.7)&32.3(0.1)&33.8(3.1)&15.5(9.1)&144&49.4 (9.9)&N& \\ 
G341.23-0.27&16 52 34.2&-44 28 36&45&6.7(0.3)&14.2(0.2)&18.8(1.1)&13.6(6.0)&115&70.3 (7.3)&N& \\ 
G342.15+0.51&16 52 28.3&-43 16 08&50&6.6(0.2)&12.4(0.1)&15.7(0.4)&16.4(0.6)&121&70.5 (2.5)&N& \\ 
G342.90-0.12&16 57 46.5&-43 04 42&151&19.2(1.1)&117.1(1.5)&316.7(3.0)&419.0(6.9)&662&2100.0 (8.1)&N&8,24 \\ 
G343.19-0.08(a)&16 58 34.9&-42 49 46&163&68.4(1.3)&146.0(0.7)&169.1(12.8)&193.9(19.8)&726&3732.1 (114.7)&N&24 \\ 
G343.40-0.40&17 00 40.4&-42 51 33&66&17.3(0.8)&34.5(0.8)&47.6(2.6)&40.2(7.5)&109&94.4 (10.1)&N& \\ 
G343.42-0.33&17 00 26.3&-42 48 03&127&36.8(0.2)&74.6(0.9)&122.4(0.9)&141.9(31.8)&167&298.7 (14.4)&N& \\ 
G343.50+0.03&16 59 10.7&-42 31 07&78&12.7(0.4)&27.4(0.5)&9.5(4.9)&31.8(1.3)&121&168.6 (13.9)&N& \\ 
G343.53-0.51(a)&17 01 32.2&-42 49 36&121&13.1(1.1)&62.4(0.3)&88.0(0.6)&75.0(10.2)&559&621.6 (17.0)&N& \\ 
G343.78-0.24&17 01 13.1&-42 27 48&204&23.3(2.1)&50.3(2.0)&39.7(7.9)&15.4(7.6)&766&2315.2 (36.2)&N& \\ 
G344.21-0.62&17 04 17.8&-42 21 09&200&111.5(0.2)&272.8(1.4)&358.0(4.3)&320.0(11.6)&645&1046.7 (351.3)&N&24 \\ 
G348.17+0.46&17 12 10.9&-38 31 59&69&14.3(1.6)&44.4(1.2)&92.0(4.3)&145.9(9.8)&559&2432.5 (135.0)&N&24 \\ 
\enddata
\end{deluxetable}

\begin{deluxetable}{lcccc}
\tablewidth{0pt}
\tablecaption{[4.5] Integrated Flux Densities: "possible" MYSO outflow candidates \label{fluxtable3}} 
\tablehead{ 
\colhead{Name} & 
\multicolumn{2}{c}{J2000 Coordinates} &
\colhead{area((\arcsec)$^{2}$)} &
\colhead{Integrated flux(mJy)} \\
\colhead{} &
\colhead{$\alpha$ ($^{\rm h}~~^{\rm m}~~^{\rm s}$)} &
\colhead{$\delta$ ($^{\circ}~~{\arcmin}~~{\arcsec}$)} &
\colhead{IRAC} &
\colhead{[4.5]} 
}
\tablecolumns{5}
\tabletypesize{\scriptsize}
\setlength{\tabcolsep}{0.05in}
\startdata
G12.20-0.03&18 12 23.6&-18 22 54&192&228.9(0.2) \\ 
G12.42+0.50&18 10 51.1&-17 55 50&1767&1404.7(4.3) \\ 
G12.68-0.18&18 13 54.7&-18 01 47&300&159.8(4.6) \\ 
G17.96+0.08&18 23 21.0&-13 15 11&117&141.9(5.0) \\ 
G19.61-0.14&18 27 16.8&-11 53 51&53&21.7(1.3) \\ 
G20.24+0.07&18 27 44.6&-11 14 54&105&81.3(1.6) \\ 
G21.24+0.19&18 29 10.2&-10 18 11&20&6.3(0.3) \\ 
G23.82+0.38&18 33 19.5&-07 55 37&85&49.5(1.9) \\ 
G24.11-0.17&18 35 52.6&-07 55 17&147&592.6(1.0) \\ 
G24.11-0.18&18 35 53.0&-07 55 23&55&50.7(0.2) \\ 
G24.33+0.14&18 35 08.1&-07 35 04&104&67.8(2.7) \\ 
G28.85-0.23&18 44 47.5&-03 44 15&53&6.6(0.1) \\ 
G29.89-0.77&18 48 37.7&-03 03 44&45&7.9(0.4) \\ 
G29.91-0.81&18 48 47.6&-03 03 31&65&20.3(0.2) \\ 
G35.83-0.20&18 57 26.9&+02 29 00&42&3.1(0.5) \\ 
G39.39-0.14&19 03 45.3&+05 40 43&124&213.0(0.2) \\ 
G40.60-0.72&19 08 03.3&+06 29 15&120&73.7(1.6) \\ 
G43.04-0.45(a)&19 11 38.9&+08 46 39&52&19.5(0.2) \\ 
G43.04-0.45(b)&19 11 39.1&+08 46 32&27&9.6(0.1) \\ 
G45.47+0.13&19 14 07.3&+11 12 16&59&121.2(0.5) \\ 
G45.50+0.12&19 14 13.0&+11 13 30&30&6.3(0.1) \\ 
G49.91+0.37&19 21 47.5&+15 14 26&32&9.4(0.2) \\ 
G54.11-0.04&19 31 40.0&+18 43 53&29&11.7(0.1) \\ 
G54.11-0.05&19 31 42.2&+18 43 45&39&5.8(0.4) \\ 
G57.61+0.02&19 38 40.8&+21 49 35&42&9.0(0.5) \\ 
G58.78+0.64&19 38 49.6&+23 08 40&48&19.2(0.1) \\ 
G58.78+0.65&19 38 49.2&+23 08 50&33&6.2(0.2) \\ 
G305.77-0.25&13 16 30.0&-62 59 09&168&15.2(0.6) \\ 
G305.80-0.24&13 16 43.4&-62 58 29&22&5.8(0.1) \\ 
G309.38-0.13(b)&13 47 20.5&-62 18 11&27&7.5(0.5) \\ 
G309.97+0.50&13 51 05.2&-61 33 20&43&3.8(0.2) \\ 
G309.99+0.51(c)&13 51 12.9&-61 32 29&14&3.8(0.1) \\ 
G310.15+0.76&13 51 59.2&-61 15 37&516&324.4(0.9) \\ 
G310.38-0.30(a)&13 56 01.8&-62 14 15&40&4.0(0.1) \\ 
G310.38-0.30(b)&13 56 01.0&-62 14 19&46&3.0(0.5) \\ 
G310.38-0.30(c)&13 55 59.9&-62 14 27&52&1.5(0.1) \\ 
G310.38-0.30(d)&13 56 00.0&-62 14 18&29&1.3(0.2) \\ 
G311.04+0.69&13 59 18.1&-61 06 33&30&1.3(0.3) \\ 
G313.71-0.19(a)&14 22 37.4&-61 08 17&94&29.1(1.2) \\ 
G313.71-0.19(b)&14 22 35.6&-61 08 41&60&7.7(0.4) \\ 
G313.76-0.86&14 25 01.3&-61 44 57&210&156.7(0.9) \\ 
G317.44-0.37&14 51 03.0&-59 49 58&63&3.0(0.7) \\ 
G320.24-0.29&15 09 55.0&-58 25 31&29&8.4(0.7) \\ 
G323.74-0.26&15 31 45.5&-56 30 50&148&778.1(1.0) \\ 
G325.52+0.42&15 39 10.6&-54 55 40&23&7.0(0.1) \\ 
G326.37+0.94&15 41 44.1&-54 00 00 &76&3.5(0.3) \\ 
G326.64+0.76&15 43 56.6&-53 58 32&39&3.3(0.2) \\ 
G326.99-0.03&15 49 07.8&-54 23 02&40&4.4(0.5) \\ 
G327.57-0.85&15 55 47.3&-54 39 09&125&373.9(2.7) \\ 
G327.65+0.13&15 52 00.5&-53 50 41&50&96.0(0.3) \\ 
G328.16+0.59&15 52 42.5&-53 09 51&251&53.0(0.7) \\ 
G328.60+0.27&15 56 15.8&-53 07 50&50&3.5(0.7) \\ 
G328.81+0.63&15 55 48.4&-52 43 06&851&1461.9(5.3) \\ 
G330.95-0.18&16 09 52.7&-51 54 56&181&225.1(0.7) \\ 
G331.12-0.46&16 11 55.3&-52 00 10&58&6.4(0.3) \\ 
G331.34-0.35&16 12 26.4&-51 46 17&94&460.4(1.6) \\ 
G331.51-0.34&16 13 11.7&-51 39 12&102&4.4(0.7) \\ 
G332.28-0.55&16 17 41.8&-51 16 04&53&8.2(0.2) \\ 
G332.29-0.09&16 15 45.2&-50 55 52&62&153.3(0.4) \\ 
G332.29-0.55&16 17 44.3&-51 15 42&26&3.2(0.7) \\ 
G332.35-0.44&16 17 31.4&-51 08 22&33&11.9(0.2) \\ 
G333.13-0.56&16 21 36.1&-50 40 49&386&38.7(4.2) \\ 
G333.32+0.10&16 19 28.9&-50 04 40&60&19.8(0.8) \\ 
G335.59-0.30&16 31 02.5&-48 44 07&176&21.8(0.6) \\ 
G336.02-0.83&16 35 09.7&-48 46 44&226&49.9(1.5) \\ 
G336.03-0.82&16 35 09.6&-48 45 55&35&10.8(0.3) \\ 
G337.40-0.40&16 38 50.4&-47 28 04&386&354.0(4.0) \\ 
G338.42-0.41&16 42 50.5&-46 42 29&256&30.1(2.5) \\ 
G338.92+0.55(a)&16 40 33.6&-45 41 44&104&125.8(0.6) \\ 
G338.92+0.55(b)&16 40 34.1&-45 42 07&230&750.6(2.3) \\ 
G339.58-0.13&16 45 59.5&-45 38 44&222&232.7(1.4) \\ 
G340.05-0.25&16 48 14.7&-45 21 52&128&27.6(2.7) \\ 
G340.06-0.23&16 48 09.7&-45 20 58&217&71.8(18.2) \\ 
G340.07-0.24&16 48 15.1&-45 20 57&23&2.6(0.1) \\ 
G340.10-0.18&16 48 07.0&-45 17 06&124&25.2(2.6) \\ 
G340.76-0.12&16 50 14.6&-44 44 38&30&3.7(0.4) \\ 
G341.22-0.26(b)&16 52 30.3&-44 28 40&109&24.9(2.3) \\ 
G342.04+0.43&16 52 27.8&-43 24 17&52&5.9(0.4) \\ 
G343.19-0.08(b)&16 58 35.8&-42 49 38&63&9.7(0.3) \\ 
G343.42-0.37&17 00 37.4&-42 49 40&17&1.9(0.1) \\ 
G343.53-0.51(b)&17 01 33.5&-42 49 50&52&13.6(0.1) \\ 
G344.22-0.57&17 04 06.4&-42 18 58&174&29.2(0.8) \\ 
G345.00-0.22(a)&17 05 11.2&-41 29 03&510&583.1(0.9) \\ 
G345.00-0.22(b)&17 05 10.9&-41 29 13&164&88.8(0.3) \\ 
G345.13-0.17&17 05 23.1&-41 21 11&78&34.4(0.2) \\ 
G346.04+0.05&17 07 19.9&-40 29 49&46&15.7(0.7) \\ 
G346.28+0.59&17 05 51.3&-39 58 43&268&324.4(1.5) \\ 
G348.18+0.48&17 12 08.0&-38 30 52&275&113.7(10.2) \\ 
G348.72-1.04&17 20 06.1&-38 57 15&36&33.0(5.0) \\ 
G348.73-1.04&17 20 06.5&-38 57 08&35&67.6(14.6) 
\enddata
\end{deluxetable}

\begin{deluxetable}{lccccccc}
\tablewidth{0pt}
\tablecaption{"Outflow-only" [3.6] and [4.5] Integrated Flux Densities \label{fluxoutflowonly}} 
\tablehead{ 
\colhead{Name} & 
\multicolumn{2}{c}{J2000 Coordinates} &
\colhead{area((\arcsec)$^{2}$)} &
\multicolumn{2}{c}{Integrated flux(mJy)}&
\colhead{[3.6]-[4.5]}\\
\colhead{} &
\colhead{$\alpha$ ($^{\rm h}~~^{\rm m}~~^{\rm s}$)} &
\colhead{$\delta$ ($^{\circ}~~{\arcmin}~~{\arcsec}$)} &
\colhead{IRAC} &
\colhead{[3.6]} &
\colhead{[4.5]} &
\colhead{} 
}
\tablecolumns{8}
\tabletypesize{\scriptsize}
\setlength{\tabcolsep}{0.05in}
\startdata
G12.91-0.26&18 14 39.5&-17 52 00&965&172.3(13.2)&385.7(23.4)&0.87 \\ 
G19.01-0.03 O-N\tablenotemark{a}&18 25 44.8&-12 22 45.76&507&66.2(4.3)&141.3(3.0)&0.82 \\ 
G19.01-0.03 O-S\tablenotemark{a}&18 25 44.8&-12 22 45.76&98&11.7(0.7)&28.5(1.3)&0.97 \\ 
G34.26+0.15&18 53 16.4&+01 15 07&3787&324.8(8.6)&838.9(15.6)&1.03 \\ 
G37.55+0.20&18 59 07.5&+04 12 31&626&13.3(3.5)&39.2(5.4)&1.17 \\ 
G345.51+0.35&17 04 24.6&-40 43 57&1407&244.4(11.4)&386.7(9.3)&0.50 \\ 
\enddata
\tablenotetext{a}{Coordinates are for the central source of the EGO
  G19.01-0.03, as given in Table~\ref{fluxtable1}.}
\end{deluxetable}

\begin{deluxetable}{lccc}
\tablewidth{0pt}
\tablecaption{Infrared EGO associations. \label{irdc_table}} 
\tablehead{ 
\colhead{Name} & 
\colhead{IRDC?} &
\colhead{Nearest IRAS PS\tablenotemark{a}} & 
\colhead{Ang. Sep.(\arcsec)}}
\tablecolumns{4}
\tabletypesize{\scriptsize}
\setlength{\tabcolsep}{0.05in}
\startdata
Table 1 &  &  & \\ 
G11.92-0.61 & Y & 18110-1854 & 62\\ 
G12.02-0.21 & Y & 18097-1835  & 137\\ 
G12.91-0.03 & Y & 18111-1746 & 200\\ 
G14.33-0.64 & Y & 18159-1648 & 15\\ 
G14.63-0.58 & Y & 18164-1631 & 34\\ 
G16.61-0.24 & Y & 18190-1435 & 133\\ 
G18.67+0.03 & N & 18220-1241 & 36\\ 
G18.89-0.47 & Y & 18242-1241 & 113\\ 
G19.01-0.03 & Y & 18228-1224 & 88\\ 
G19.88-0.53 & Y & 18264-1152 & 6\\ 
G22.04+0.22 & Y & 18278-0936 & 5\\ 
G23.01-0.41 & N & 18318-0901 & 130\\ 
G23.96-0.11 & N & 18326-0802 & 113\\ 
G24.00-0.10 & Y & 18326-0802 & 50\\ 
G24.17-0.02 & Y & 18326-0751 & 64\\ 
G24.94+0.07 & N & 18337-0707 & 78\\ 
G25.27-0.43 & Y & 18362-0703 & 29\\ 
G27.97-0.47 & Y & 18412-0440 & 116\\ 
G28.83-0.25 & Y & 18421-0348 & 35\\ 
G34.41+0.24 & Y & 18507+0121 & 31\\ 
G35.03+0.35 & N & 18515+0157 & 58\\ 
G35.04-0.47 & Y & 18545+0134 & 135\\ 
G35.13-0.74 & N & 18553+0133 & 215\\ 
G35.15+0.80 & N & 18500+0216 & 24\\ 
G35.20-0.74 & N & 18556+0136 & 4\\ 
G35.68-0.18 & Y & 18547+0215 & 246\\ 
G35.79-0.17 & Y & 18547+0223 & 10\\ 
G36.01-0.20 & Y & 18551+0238 & 194\\ 
G37.48-0.10 & N & 18574+0355 & 191\\ 
G39.10+0.49 & N & 18585+0538 & 5\\ 
G40.28-0.27 & Y & 19034+0618 & 70\\ 
G44.01-0.03 & N & 19094+0944 & 79\\ 
G45.47+0.05 & Y & 19120+1103 & 59\\ 
G49.27-0.34 & Y & 19207+1410 & 221\\ 
G54.45+1.02 & N & 19262+1925 & 12\\ 
G56.13+0.22 & N & 19326+2030 & 19\\ 
G58.09-0.34 & Y & 19388+2156 & 23\\ 
G59.79+0.63 & Y & 19389+2354 & 10\\ 
G298.26+0.74 & N & 12091-6129 & 3\\ 
G298.89+0.37 & Y & 12140-6157 & 76\\ 
G298.90+0.36 & N & 12140-6157 & 19\\ 
G305.48-0.10 & N & 13108-6233 & 191\\ 
G305.52+0.76 & Y & 13102-6143 & 12\\ 
G305.57-0.34 & Y & 13113-6249 & 109\\ 
G305.62-0.34 & N & 13119-6249 & 7\\ 
G305.82-0.11 & Y & 13128-6236 & 291\\ 
G305.89+0.02 & Y & 13140-6226 & 27\\ 
G309.15-0.35 & Y & 13421-6217 & 125\\ 
G309.38-0.13(a) & Y & 13438-6203 & 32\\ 
G309.99+0.51(a) & Y & 13475-6115 & 129\\ 
G310.08-0.23 & Y & 13497-6159 & 31\\ 
G312.11+0.26 & Y & 14050-6056 & 171\\ 
G317.42-0.67 & Y & 14481-5953 & 5\\ 
G317.46-0.40(b) & Y & 14480-5941 & 313\\ 
G317.87-0.15 & Y & 14488-5915 & 276\\ 
G317.88-0.25 & Y & 14500-5918 & 78\\ 
G321.94-0.01 & Y & 15158-5707 & 34\\ 
G324.19+0.41 & N & 15277-5532 & 39\\ 
G324.72+0.34 & Y & 15310-5517 & 13\\ 
G326.27-0.49 & Y & 15432-5501 & 7\\ 
G326.31+0.90 & Y & 15375-5352 & 168\\ 
G326.78-0.24 & N & 15450-5431 & 7\\ 
G326.79+0.38 & Y & 15425-5401 & 2\\ 
G326.86-0.67 & Y & 15475-5449 & 104\\ 
G326.97-0.03 & Y & 15453-5416 & 164\\ 
G327.12+0.51 & N & 15437-5343 & 4\\ 
G327.39+0.20 & N & 15464-5348* & 4\\ 
G327.40+0.44 & Y & 15454-5335 & 29\\ 
G328.14-0.43 & Y & 15529-5350 & 94\\ 
G329.18-0.31 & Y & 15579-5303 & 8\\ 
G329.47+0.52 & Y & 15557-5215 & 53\\ 
G329.61+0.11 & Y & 15584-5230 & 208\\ 
G332.35-0.12 & Y & 16122-5047 & 18\\ 
G332.56-0.15 & Y & 16132-5039 & 97\\ 
G332.81-0.70 & N & 16170-5053 & 48\\ 
G332.94-0.69 & Y & 16175-5046 & 22\\ 
G332.96-0.68 & Y & 16175-5045 & 3\\ 
G333.18-0.09 & Y & 16159-5012 & 75\\ 
G335.59-0.29 & Y & 16272-4837 & 21\\ 
G337.30-0.87 & Y & 16366-4746 & 82\\ 
G338.39-0.40 & Y & 16390-4637 & 25\\ 
G339.95-0.54 & N & 16455-4531 & 90\\ 
G340.97-1.02 & Y & 16513-4504 & 48\\ 
G341.24-0.27 & Y & 16487-4423 & 170\\ 
G341.73-0.97 & Y & 16537-4426 & 3\\ 
G341.99-0.10 & Y & 16510-4341 & 55\\ 
G343.12-0.06 & Y & 16547-4247 & 4\\ 
G343.50-0.47 & N & 16579-4245 & 120\\ 
G343.72-0.18(a) & Y & 16572-4221 & 137\\ 
G343.72-0.18(b) & Y & 16574-4225 & 133\\ 
G344.58-0.02 & N & 16594-4137 & 9\\ 
G345.72+0.82 & Y & 16596-4012 & 0.4\\ 
G345.99-0.02 & N & 17039-4030 & 14\\ 
G347.08-0.40 & Y & 17089-3951 & 9\\ 
G348.55-0.98 & N & 17158-3901 & 64\\ 
G348.58-0.92 & Y & 17157-3855 & 129\\ 
G349.15-0.98 & Y & 17176-3831 & 0.8\\ 
  & 70/97 & avg./median & 75/50\\ 
\hline
Table 2 &  &  & \\ 
G10.29-0.13 & Y & 18060-2005 & 134\\ 
G10.34-0.14 & Y & 18060-2005 & 111\\ 
G16.59-0.05 & Y & 18182-1433 & 19\\ 
G19.36-0.03 & Y & 18236-1205 & 24\\ 
G19.61-0.12 & N & 18244-1155 & 43\\ 
G25.38-0.15 & Y & 18355-0650 & 110\\ 
G28.28-0.36 & N & 18416-0420 & 81\\ 
G34.39+0.22 & Y & 18507+0121 & 52\\ 
G48.66-0.30  & Y & 19199+1347 & 463\\ 
G49.42+0.33 & N & 19186+1440 & 34\\ 
G309.90+0.23 & Y & 13475-6135 & 54\\ 
G317.46-0.40(a) & Y & 14480-5941 & 319\\ 
G318.04+0.09 & Y & 14498-5856 & 24\\ 
G318.05+0.09 & N & 14498-5856 & 9\\ 
G320.23-0.28 & Y & 15061-5814 & 82\\ 
G324.17+0.44 & N & 15277-5532 & 102\\ 
G326.32-0.39 & N & 15428-5453 & 213\\ 
G326.47+0.70 & Y & 15394-5358 & 27\\ 
G326.48+0.70 & Y & 15394-5358 & 20\\ 
G326.61+0.80(b) & Y & 15395-5348 & 89\\ 
G328.25-0.53 & N & 15541-5349 & 3\\ 
G329.03-0.20 & Y & 15566-5304 & 2\\ 
G329.41-0.46 & N & 15596-5301 & 11\\ 
G329.47+0.50 & Y & 15557-5215 & 48\\ 
G330.88-0.37 & Y & 16065-5158 & 20\\ 
G331.13-0.24 & N & 16071-5142 & 28\\ 
G332.12+0.94 & Y & 16067-5010 & 10\\ 
G332.60-0.17 & Y & 16136-5038 & 69\\ 
G332.73-0.62 & Y & 16158-5055 & 262\\ 
G333.47-0.16 & N & 16175-5002 & 14\\ 
G335.06-0.43 & Y & 16256-4905 & 17\\ 
G335.79+0.18 & Y & 16259-4805 & 224\\ 
G337.91-0.48 & N & 16374-4701 & 42\\ 
G341.22-0.26(a) & Y & 16487-4423 & 121\\ 
G342.48+0.18 & Y & 16515-4308 & 51\\ 
G344.23-0.57 & Y & 17006-4215 & 94\\ 
  & 25/36 & avg./median & 84/50\\ 
\hline
Table 3 &  &  & \\ 
G11.11-0.11 & Y & 18073-1923 & 104\\ 
G16.58-0.08 & Y & 18182-1433 & 125\\ 
G24.63+0.15 & Y & 18331-0717 & 281\\ 
G29.84-0.47 & Y & 18446-0303 & 190\\ 
G29.96-0.79 & Y & 18461-0304 & 35\\ 
G34.28+0.18 & Y & 18507+0110 & 170\\ 
G40.28-0.22 & Y & 19031+0621 & 73\\ 
G45.80-0.36 & N & 19141+1110 & 10\\ 
G49.07-0.33 & Y & 19205+1403 & 179\\ 
G49.27-0.32 & N & 19207+1410 & 252\\ 
G50.36-0.42 & Y & 19234+1510 & 144\\ 
G53.92-0.07 & N & 19291+1826 & 20\\ 
G54.11-0.08 & N & 19294+1836 & 89\\ 
G54.45+1.01 & N & 19262+1925 & 9\\ 
G58.79+0.63 & N & 19366+2301 & 105\\ 
G62.70-0.51 & N & 19497+2549 & 14\\ 
G304.89+0.64 & Y & 13050-6154 & 6\\ 
G309.91+0.32 & Y & 13475-6129 & 69\\ 
G309.97+0.59 & Y & 13475-6115 & 169\\ 
G309.99+0.51(b) & Y & 13475-6115 & 142\\ 
G311.51-0.45 & Y & 14023-6150A & 101\\ 
G324.11+0.44 & Y & 15272-5533 & 2\\ 
G326.36+0.88 & Y & 15375-5352 & 308\\ 
G326.41+0.93 & Y & 15384-5348 & 160\\ 
G326.57+0.20 & N & 15422-5418 & 112\\ 
G326.61+0.80(a) & Y & 15395-5348 & 102\\ 
G326.61+0.80(c) & Y & 15395-5348 & 95\\ 
G326.65+0.75 & Y & 15402-5349 & 12\\ 
G326.80+0.51 & Y & 15415-5356 & 217\\ 
G326.92-0.31 & Y & 15462-5428 & 110\\ 
G327.30-0.58 & Y & 15492-5426 & 101\\ 
G327.72-0.38 & N & 15508-5403 & 84\\ 
G327.86+0.19 & N & 15489-5333 & 121\\ 
G327.89+0.15 & Y & 15489-5333 & 228\\ 
G328.55+0.27 & Y & 15522-5301 & 12\\ 
G329.07-0.31(a) & Y & 15573-5307 & 17\\ 
G329.07-0.31(b) & Y & 15573-5307 & 1\\ 
G329.16-0.29 & Y & 15579-5303 & 120\\ 
G329.18-0.30 & Y & 15579-5303 & 30\\ 
G331.08-0.47 & Y & 16076-5154 & 161\\ 
G331.37-0.40 & Y & 16089-5137 & 154\\ 
G331.62+0.53 & N & 16061-5048 & 21\\ 
G331.71+0.58 & Y & 16062-5041 & 70\\ 
G331.71+0.60 & Y & 16062-5041 & 3\\ 
G332.28-0.07 & Y & 16119-5048 & 100\\ 
G332.33-0.12 & Y & 16122-5047 & 61\\ 
G332.36+0.60 & Y & 16093-5015 & 38\\ 
G332.47-0.52 & Y & 16147-5100 & 47\\ 
G332.58+0.15 & Y & 16122-5028 & 167\\ 
G332.59+0.04(a) & Y & 16128-5033 & 185\\ 
G332.59+0.04(b) & Y & 16128-5033 & 193\\ 
G332.91-0.55 & Y & 16168-5044 & 156\\ 
G333.08-0.56 & Y & 16172-5032 & 278\\ 
G334.04+0.35 & N & 16178-4916 & 7\\ 
G334.25+0.07 & Y & 16200-4919 & 48\\ 
G335.43-0.24 & Y & 16264-4841 & 77\\ 
G336.87+0.29 & Y & 16301-4718 & 138\\ 
G336.96-0.98 & Y & 16358-4804 & 32\\ 
G337.16-0.39 & Y & 16340-4732 & 22\\ 
G338.32-0.41 & Y & 16389-4639 & 129\\ 
G340.75-1.00 & Y & 16506-4512 & 161\\ 
G340.77-0.12 & Y & 16465-4437 & 126\\ 
G340.78-0.10 & N & 16465-4437 & 52\\ 
G341.20-0.26 & Y & 16487-4423 & 110\\ 
G341.23-0.27 & Y & 16487-4423 & 141\\ 
G342.15+0.51 & N & 16489-4318 & 453\\ 
G342.90-0.12 & N & 16541-4259 & 41\\ 
G343.19-0.08(a) & Y & 16550-4245 & 10\\ 
G343.40-0.40 & Y & 16566-4249 & 345\\ 
G343.42-0.33 & Y & 16573-4240 & 342\\ 
G343.50+0.03 & Y & 16558-4228 & 158\\ 
G343.53-0.51(a) & Y & 16579-4245 & 45\\ 
G343.78-0.24 & Y & 16576-4223 & 8\\ 
G344.21-0.62 & N & 17006-4215 & 95\\ 
G348.17+0.46 & N & 17089-3834 & 383\\ 
 & 58/75 & avg./median & 116/102\\ 
\hline
Table 4 &  &  & \\ 
G12.2-0.03 & N & 18094-1823 & 22\\ 
G12.42+0.50 & N & 18079-1756 & 12\\ 
G12.68-0.18 & N & 18112-1801 & 206\\ 
G17.96+0.08 & N & 18205-1316 & 12\\ 
G19.61-0.14 & N & 18244-1155 & 13\\ 
G20.24+0.07 & N & 18249-1116 & 10\\ 
G21.24+0.19 & Y & 18263-1020 & 19\\ 
G23.82+0.38 & N & 18305-0758 & 22\\ 
G24.11-0.17 & Y & 18334-0757 & 216\\ 
G24.11-0.18 & Y & 18334-0757 & 210\\ 
G24.33+0.14 & Y & 18324-0737 & 5\\ 
G28.85-0.23 & N & 18421-0348 & 78\\ 
G29.89-0.77 & Y & 18460-0307 & 25\\ 
G29.91-0.81 & N & 18460-0307 & 128\\ 
G35.83-0.20 & Y & 18549+0226 & 83\\ 
G39.39-0.14 & N & 19012+0536 & 4\\ 
G40.60-0.72 & N & 19056+0624 & 8\\ 
G43.04-0.45(a) & N & 19092+0841 & 24\\ 
G43.04-0.45(b) & N & 19092+0841 & 25\\ 
G45.47+0.13 & N & 19117+1107 & 21\\ 
G45.50+0.12 & N & 19117+1107 & 91\\ 
G49.91+0.37 & N & 19195+1508 & 5\\ 
G54.11-0.04 & N & 19294+1836 & 89\\ 
G54.11-0.05 & N & 19294+1836 & 73\\ 
G57.61+0.02 & N & 19365+2142 & 18\\ 
G58.78+0.64 & N & 19366+2301 & 23\\ 
G58.78+0.65 & N & 19366+2301 & 21\\ 
G305.77-0.25 & N & 13134-6242 & 91\\ 
G305.80-0.24 & N & 13134-6242 & 17\\ 
G309.38-0.13(b) & Y & 13438-6203 & 32\\ 
G309.97+0.50 & Y & 13475-6115 & 184\\ 
G309.99+0.51(c) & Y & 13475-6115 & 149\\ 
G310.15+0.76 & N & 13484-6100 & 14\\ 
G310.38-0.30(a) & Y & 13524-6159 & 9\\ 
G310.38-0.30(b) & Y & 13524-6159 & 13\\ 
G310.38-0.30(c) & Y & 13524-6159 & 24\\ 
G310.38-0.30(d) & Y & 13524-6159 & 17\\ 
G311.04+0.69 & Y & 13558-6051 & 57\\ 
G313.71-0.19(a) & Y & 14188-6054 & 37\\ 
G313.71-0.19(b) & Y & 14188-6054 & 33\\ 
G313.76-0.86 & Y & 14212-6131 & 11\\ 
G317.44-0.37 & Y & 14467-5939 & 242\\ 
G320.24-0.29 & Y & 15061-5814 & 64\\ 
G323.74-0.26 & N & 15278-5620 & 30\\ 
G325.52+0.42 & N & 15353-5445 & 12\\ 
G326.37+0.94 & Y & 15375-5352 & 233\\ 
G326.64+0.76 & Y & 15402-5349 & 52\\ 
G326.99-0.03 & Y & 15453-5416 & 151\\ 
G327.57-0.85 & Y & 15519-5430 & 29\\ 
G327.65+0.13 & N & 15481-5341 & 2\\ 
G328.16+0.59 & N & 15488-5300 & 1\\ 
G328.60+0.27 & Y & 15522-5301 & 163\\ 
G328.81+0.63 & Y & 15520-5234 & 4\\ 
G330.95-0.18 & Y & 16060-5146 & 35\\ 
G331.12-0.46 & Y & 16082-5150 & 158\\ 
G331.34-0.35 & N & 16085-5138 & 41\\ 
G331.51-0.34 & Y & 16093-5128 & 168\\ 
G332.28-0.55 & Y & 16141-5107 & 159\\ 
G332.29-0.09 & N & 16119-5048 & 10\\ 
G332.29-0.55 & Y & 16141-5107 & 129\\ 
G332.35-0.44 & Y & 16137-5100 & 10\\ 
G333.13-0.56 & N & 16178-5029 & 237\\ 
G333.32+0.10 & Y & 16157-4957 & 2\\ 
G335.59-0.30 & Y & 16272-4837 & 63\\ 
G336.02-0.83 & N & 16313-4840 & 33\\ 
G336.03-0.82 & N & 16313-4840 & 27\\ 
G337.40-0.40 & Y & 16351-4722 & 8\\ 
G338.42-0.41 & Y & 16390-4637 & 95\\ 
G338.92+0.55(a) & N & 16368-4538 & 171\\ 
G338.92+0.55(b) & N & 16368-4538 & 151\\ 
G339.58-0.13 & Y & 16424-4531 & 139\\ 
G340.05-0.25 & Y & 16445-4516 & 34\\ 
G340.06-0.23 & Y & 16445-4516 & 43\\ 
G340.07-0.24 & Y & 16445-4516 & 51\\ 
G340.10-0.18 & N & 16442-4511 & 169\\ 
G340.76-0.12 & Y & 16465-4437 & 130\\ 
G341.22-0.26(b) & Y & 16487-4423 & 102\\ 
G342.04+0.43 & N & 16489-4318 & 77\\ 
G343.19-0.08(b) & Y & 16550-4245 & 22\\ 
G343.42-0.37 & Y & 16573-4240 & 331\\ 
G343.53-0.51(b) & Y & 16579-4245 & 51\\ 
G344.22-0.57 & Y & 17006-4215 & 89\\ 
G345.00-0.22(a) & Y & 17016-4124 & 16\\ 
G345.00-0.22(b) & Y & 17016-4124 & 15\\ 
G345.13-0.17 & N & 17020-4117 & 129\\ 
G346.04+0.05 & N & 17038-4026 & 60\\ 
G346.28+0.59 & N & 17023-3954 & 16\\ 
G348.18+0.48 & N & 17089-3820 & 404\\ 
G348.72-1.04 & Y & 17167-3854 & 40\\ 
G348.73-1.04 & Y & 17167-3854 & 39\\ 
 & 49/90 & avg./median & 73/36\\ 
\hline
Outflow-only (Table 5) &  &  & \\ 
G12.91-0.26 & N & 18117-1753 & 8\\ 
G34.26+0.15 & N & 18507+0110 & 56\\ 
G37.55+0.20 & N & 18566+0408 & 39\\ 
G345.51+0.35 & N & 17008-4040 & 55\\ 
  & 0/4  & avg./median & 40/47\\ 
\enddata
\tablenotetext{a}{B1950 coordinates}
\end{deluxetable}

\begin{deluxetable}{lccccc}
\tablewidth{0pt}
\tablecaption{6.7GHz CH$_{3}$OH maser associations of EGOs \label{maser_table}} 
\tablehead{ 
\colhead{Name} & 
\colhead{Maser Survey\tablenotemark{a}} &
\colhead{Walsh IRAS target\tablenotemark{b}} & 
\colhead{Maser?\tablenotemark{a}}&
\colhead{Offset(\pp)\tablenotemark{c}} &
\colhead{Distance(kpc)}\tablenotemark{d}}
\tablecolumns{5}
\tabletypesize{\scriptsize}
\setlength{\tabcolsep}{0.05in}
\startdata
Table 1 &  &  &  & &\\ 
G11.92-0.61 & W  & 18110-1854 & N & & \\ 
G14.33-0.64 & W  & 18159-1648 & N & & \\ 
G28.83-0.25 & W  & 18421-0348 & Y & 3 & \\ 
G298.26+0.74 & W  & 12091-6129 & Y & $<$1 & \\ 
G312.11+0.26 & W  & 14050-6056 & Y & 1 & \\ 
G326.27-0.49 & E  &  & N & & \\ 
G326.78-0.24 & E  &  & N & & \\ 
G326.79+0.38 & E  &  & N & & \\ 
G326.86-0.67\tablenotemark{e} & E  &  & Y & 5 & 3.7\\ 
G326.97-0.03 & E  &  & N & & \\ 
G327.12+0.51 & W,E & 15437-5343 & Y (W,E) & 1 & 5.5\\ 
G327.39+0.20 & E  &  & Y & $<$1 & 5.3\\ 
G327.40+0.44 & E  &  & Y & 5 & 5.2\\ 
G328.14-0.43 & E  &  & N & &  \\ 
G329.18-0.31 & E  &  & Y & 3 & 3.6\\ 
G329.47+0.52 & W,E & 15557-5215 & Y (W) & $<$1 & 4.5\\ 
G329.61+0.11 & E  &  & Y & $<$1 & 3.9\\ 
G332.35-0.12 & C,E &  & N & & \\ 
G332.56-0.15 & C,W,E & 16132-5039 & Y (C,W,E) & 2 & 3.5\\ 
G332.94-0.69\tablenotemark{e} & E  &  & Y & 1 & 3.6\\ 
G333.18-0.09 & C,E & 16159-5012 & Y(C,E, W could not locate) & 1 & 5\\ 
G335.59-0.29 & C,W,E & 16272-4837 & Y (C,W) & 2,4,17 & \\ 
G339.95-0.54 & W  & 16455-4531 & Y & 1 & \\ 
G343.12-0.06 & W  & 16547-4247 & N & & \\ 
G344.58-0.02 & W  & 16594-4137 & Y & $<$1 & \\ 
\hline
Table 2 &  &  &  & &\\ 
G10.29-0.13 & W  & 18060-2005 & Y & 2 & \\ 
G10.34-0.14 & W  & 18060-2005 & Y & $<$1 & \\ 
G16.59-0.05 & W  & 18182-1433 & Y & $<$1 & \\ 
G19.36-0.03 & W  & 18236-1205 & Y & 4 & \\ 
G19.61-0.12 & W  & 18244-1155 & Y & 5 & \\ 
G28.28-0.36 & W  & 18416-0420 & Y & 1 & \\ 
G320.23-0.28 & W  & 15061-5814 & Y & 6 & \\ 
G326.32-0.39 & E  &  & N & & \\ 
G326.47+0.70\tablenotemark{e} & E  &  & Y? & 11 & 2.6\\ 
G326.48+0.70\tablenotemark{e} & E  &  & Y? & 8 & 2.6\\ 
G328.25-0.53 & E  &  & Y & 1 & 2.6\\ 
G329.03-0.20 & E  &  & Y & 7,19 & 2.6,2.9\\ 
G329.41-0.46 & W,E & 15596-5301 & Y(E) & 3 & 4.2\\ 
G329.47+0.50 & W,E & 15557-5215 & Y & 3 & 4.5\\ 
G330.88-0.37 & W,E & 16065-5158 & N & & \\ 
G331.13-0.24 & W,E & 16071-5142 & Y (W,E) & 3 & 5.2\\ 
G332.60-0.17 & C,E &  & Y(C,E) & $<$1 & 3.5\\ 
G332.73-0.62 & W  & 16158-5055 & Y & 1 & \\ 
G333.47-0.16 & C,W,E & 16175-5002 & Y (C,W,E) & 1 & 3\\ 
G335.06-0.43 & E  &  & Y & $<$1 & 3.4\\ 
G335.79+0.18 & C  &  & Y & 5 & \\ 
G337.91-0.48 & W  & 16374-4701 & N & & \\ 
G344.23-0.57 & W  & 17006-4215 & Y & 7,8 & \\ 
\hline
Table 3 &  &  &  & &\\ 
G16.58-0.08 & W  & 18182-1433 & N & &\\ 
G326.57+0.20 & E  &  & N & &\\ 
G326.80+0.51 & E  &  & N & &\\ 
G326.92-0.31 & E  &  & N & &\\ 
G327.72-0.38 & E  &  & N & &\\ 
G327.86+0.19 & E  &  & N & &\\ 
G327.89+0.15 & E  &  & N & &\\ 
G328.55+0.27 & E  &  & N & &\\ 
G329.07-0.31(a) & E  &  & N & &\\ 
G329.07-0.31(b) & E  &  & Y & $<$1 & 3\\ 
G329.16-0.29 & E  &  & N & &\\ 
G329.18-0.30 & E  &  & N & &\\ 
G331.08-0.47 & E  &  & N & &\\ 
G331.37-0.40 & W,E & 16085-5138 & N & &\\ 
G331.62+0.53 & E  &  & N & &\\ 
G332.28-0.07 & C,E &  & N & &\\ 
G332.33-0.12 & C,E &  & N & &\\ 
G332.47-0.52 & E  &  & N & &\\ 
G332.58+0.15 & C,E &  & N & &\\ 
G332.59+0.04(a) & C,E &  & N & &\\ 
G332.59+0.04(b) & C,E &  & N & &\\ 
G334.04+0.35 & E  &  & N & &\\ 
G334.25+0.07 & C,E &  & N & &\\ 
G340.75-1.00 & W  & 16506-4512 & N & &\\ 
G343.19-0.08(a) & W  & 16547-4247 & N & &\\ 
G344.21-0.62 & W  & 17006-4215 & N & &\\ 
\hline
Table 4 &  &  &  & &\\ 
G19.61-0.14 & W  & 18244-1155 & N & &\\ 
G28.85-0.23 & W  & 18421-0348 & Y & 2 &\\ 
G305.77-0.25 & W  & 13134-6242 & N && \\ 
G305.80-0.24 & W  & 13134-6242 & N & &\\ 
G309.97+0.50 & W  & 13471-6120 & N & &\\ 
G313.76-0.86 & W  & 14212-6131 & Y & 3,25 &\\ 
G320.24-0.29 & W  & 15061-5814 & N & &\\ 
G323.74-0.26 & W  & 15278-5620 & Y & 0.6,0.8,3 & \\ 
G325.52+0.42 & E  &  & N & &\\ 
G326.99-0.03 & E  &  & N & &\\ 
G327.65+0.13 & E  &  & N & &\\ 
G328.60+0.27 & E  &  & N & &\\ 
G328.81+0.63 & W,E & 15520-5234 & Y (W,E) & $<$1 & 3\\ 
G330.95-0.18 & C,W,E & 16060-5146 & Y (W,E) & 3 & 5.3\\ 
G331.12-0.46 & E  &  & N & &\\ 
G331.34-0.35 & W,E & 16085-5138 & Y(W,E) & 1 & 4.3\\ 
G331.51-0.34 & E  &  & N & & \\ 
G332.29-0.09 & C,E &  & Y(E) & 3 & 3.2\\ 
G332.35-0.44 & E  &  & Y & 2 & 3.6\\ 
G333.13-0.56\tablenotemark{e} & E  &  & Y & 10 & 3.8\\ 
G333.32+0.10 & C,E &  & Y(C,E) & 2 & 3.1\\ 
G335.59-0.30 & W  & 16272-4837 & N & &\\ 
G336.02-0.83 & W  & 16313-4840 & Y & 5 & \\ 
G336.03-0.82 & W  & 16313-4840 & N & & \\ 
G337.40-0.40 & W  & 16351-4722 & Y & 3,4 & \\ 
G339.58-0.13 & C  & 16424-4531 and 16421-4532 & Y & 8 & \\ 
G340.05-0.25 & W  & 16445-4516 & Y & 12 & \\ 
G340.06-0.23 & W  & 16445-4516 & N & &\\ 
G340.07-0.24 & W  & 16445-4516 & N & &\\ 
G343.19-0.08(b) & W  & 16547-4247 & N & &\\ 
G344.22-0.57 & W  & 17006-4215 & N & &\\ 
G345.00-0.22(a) & W  & 17016-4124 & Y & 4,5 &\\ 
G345.00-0.22(b) & W  & 17016-4124 & N & &\\ 
G348.72-1.04 & W  & 17167-3854 & N & &\\ 
G348.73-1.04 & W  & 17167-3854 & Y & $<$1 &\\ 
\hline 
Outflow-only (Table 5) &  &  &  & &\\ 
G12.91-0.26 & W  & 18117-1753 & Y & $<$1 & \\ 
G345.51+0.35 & W  & 17008-4040 & Y & 32 & \\ 
\enddata
\tablenotetext{a}{C=\citet{Caswell96}, W=\citet{Walsh98}, E=\citet{Ellingsen06}}
\tablenotetext{b}{B1950 coordinates}
\tablenotetext{c}{Nominal separation from EGO positions listed in
  Tables~\ref{fluxtable1}-\ref{fluxoutflowonly}, rounded to the
  nearest arcsecond.  Separations less than 1\pp\/ are denoted $<$1.
  Where multiple offsets are listed, multiple spatially distinct maser
spots are reported.}
\tablenotetext{d}{Distances are the near kinematic distances reported in
  \citet{Ellingsen06}.  The association of EGOs with IRDCs supports the
  adoption of the near kinematic distance.}
\tablenotetext{e}{Source outside
  the nominal area of \citet{Ellingsen96} and \citet{Ellingsen06} of 325$\degr
  <$ \emph{l} $<$ 335$\degr$,
  b=$\pm$0.53$\degr$ and not included in the calculation of the fraction of
  EGOs within the survey coverage that have associated 6.7 GHz \meth\/ masers.} 
\end{deluxetable}

\clearpage

\end{document}